\documentclass[a4paper]{tMPH2e}

\usepackage{ulem}
\usepackage{color}
\usepackage{pdflscape}

\newcommand{\ea}{{\it et al. }}

\newcommand{\LJ}{{\rm LJ}}

\newcommand{\vr}{{\bf r}}

\newcommand{\kB}{k_{\rm B}}

\newcommand{\w}{\rm H_2O}
\newcommand{\el}{\rm NaCl}

\newcommand{\vdW}{\rm vdW}
\newcommand{\Coul}{\rm Coul}

\newcommand{\ind}{\rm ind}

\newcommand{\qSpring}{\rm qSpring}

\begin{document}
\begin{center}


{\large {\bf  Recent progress in molecular simulation of aqueous electrolytes: \\
Force fields, chemical potentials and solubility}}

\bigskip

\bigskip

Ivo NEZBEDA$^{a,b}$,
Filip MOU\v{C}KA$^a$
and
William R. SMITH$^{c,d,}\footnote{%
Corresponding author. Email: bilsmith@uoguelph.ca}$
\bigskip

 $^{a}$ Faculty of Science, J. E. Purkinje University,
 400\,96 \'{U}st\'{i} nad Labem,\\[0pt] Czech Republic

$^{b}$Institute of Chemical Process Fundamentals, Academy of Sciences,
 16502 Prague 6, Czech Republic

 $^{c}$ Dept. of Mathematics and Statistics, University of Guelph,
 Guelph ON, N1G 2W1, Canada

 $^{d}$ Faculty of Science, University of Ontario Institute of Technology, Oshawa ON, L1H 7K4, Canada

\end{center}

\bigskip
\begin{abstract}
Although aqueous electrolytes are among the most important solutions, the molecular simulation of their intertwined properties of chemical potentials, solubility and activity coefficients has remained a challenging problem, and has attracted considerable recent interest.  In this perspectives review, we focus on the simplest case of aqueous sodium chloride at ambient conditions and discuss the two main factors that have impeded progress.  The first is lack of consensus with respect to the appropriate methodology for force field (FF) development.  We examine how most commonly used FFs have been developed, and emphasize the importance of distinguishing between ``Training Set Properties" used to fit the FF parameters, and ``Test Set Properties", which are pure predictions of additional properties.  The second is disagreement among solubility results obtained, even using identical FFs and thermodynamic conditions.  Solubility calculations have been approached using both thermodynamic--based methods and direct molecular dynamics--based methods implementing coexisting solution and solid phases.  Although convergence has been very recently achieved among results based on the former approach, there is as yet no
general agreement with simulation results based on the latter methodology.
We also propose a new method to directly calculate the electrolyte standard chemical potential in the Henry--Law ideality model.
We conclude by making recommendations for calculating solubility, chemical potentials and activity coefficients, and outline a potential path for future progress.
\end{abstract}

\vspace*{8cm}

\noindent{Preprint, accepted for publication in Molecular Physics 08 March 2016}

\clearpage

\section{Introduction}

Water is the most important fluid in nature, and aqueous electrolyte solutions are ubiquitous components of environmental, geochemical, industrial and biological systems.
Although biological processes generally take place at ambient conditions and at low to moderate concentration (we herein use the molality $m$ as the concentration measure), other situations involve concentrations up to the solubility limit; moreover, higher temperatures and pressures are often important in many geochemical and industrial systems.
The ability to understand and predict the properties of aqueous electrolyte solutions over a wide range of thermodynamic conditions is thus of considerable interest and importance. \\

Although empirically--based macroscopic models have been developed and used for this purpose many years (see, {\it e.g.}, \cite{Anderko2002}), in order to be capable of quantitative predictions, they require extensive experimental data to determine model parameters
for their implementation.
Most models and their parameters have been developed at or near temperature $T=298.15$ K and atmospheric pressure, $P$ ({\it i.e.}, at ambient conditions).
Many empirical models also require the assumption of the presence of species molecular complexes in addition to the solution components themselves. The net result is that macroscopic models entail many parameters and require correspondingly large experimental data sets to determine them. Although empirical models for many commonly occurring systems have been developed over the years, they cannot be reliably used outside the range of thermodynamic conditions under which they have been developed.  More importantly, such models are unavailable for systems involving previously unconsidered solution species and previously unencountered ranges of thermodynamic conditions.\\

During the period 1960--2000, an enormous effort was invested in the development of the molecular theory of electrolytes, see, {\it e.g.}, \cite{Barthel1998,Lee2008}. However, one may say that this effort produced few useful results, due to several factors.  Aqueous solutions require a tractable and accurate molecular model of water, which was not available that time for use in integral equation theories. These hence primarily used models that considered water only implicitly (primitive models of electrolytes, an approach also used in molecular simulations, due to the limitations of then available computational capabilities.  Also, the long range character of Coulombic interactions limits the capability of perturbation theories, another approach prevalent during that period.   Due to these difficulties, theoretical development in the field of aqueous electrolytes virtually ceased about two decades ago. \\

Vastly increased computer capabilities evolving since that time have led to their increased use in directly simulating material properties from molecular--level models bypassing the inherent approximations of theoretical approaches. In contrast to macroscopic approaches, these methodologies require relatively small model parameter sets to predict virtually all thermodynamic and transport quantities of interest.  Two types of approaches have become widely used tools: molecular dynamics (MD), which generates a sequence of molecular configurations over time steps, and Monte Carlo (MC), which generates a Markov chain sampling the configurational space \cite{AllenTildesley1987,Frenkel2002}.  It is important to emphasize that molecular simulations are useful not only to predict properties directly, but also to generate pseudo--experimental data that can be used to fit the parameters of macroscopic thermodynamic models, whose implementations require orders of magnitude less computer time to implement within process simulation software.\\

The use of MD and MC molecular simulation requires as input the a priori postulation of a {\it molecular interaction model} (MIM) of a given mathematical form to evaluate the molecular--level forces or energies; its realization with a specific set of parameter values we then refer to as a {\it force field} (FF).  A goal is to develop FFs that are {\it transferable}, in the sense that they can be used both for mixtures and for properties and thermodynamic conditions other than those involving the pure species for which the FF has been originally developed.
An alternative {\it ab--initio} approach iteratively calculates the forces by means of an approximate quantum mechanical approach (typically some level of density functional theory \cite{Marx2009} combined with MD). Although this approach promises the ultimate possibility of parameter--free predictions, the level of accuracy required in the quantum mechanical calculations and the resultant computational time involved mean that such calculations are at present only feasible for relatively small and simple systems ({\it e.g.}~\cite{Gomez2013}).  We thus focus on the use of the classical FF approach in this paper. \\

MD and MC techniques currently provide useful property predictions and microscopic insight for many fluid systems, including aqueous solutions \cite{Theodoru2010,Auffinger2007,Ungerer2007,Chen2007,Maginn2010,Meunier2011,Dror2012}.
However, the ability of these techniques to quantitatively predict the range of aqueous electrolyte solution properties in agreement with experiment is far from satisfactory and  continues to be a field of interest, even in the relatively simple case of aqueous solutions of simple electrolytes such as alkali halides at ambient conditions.
Progress has been impeded primarily by the lack of consensus concerning the following three factors:
\begin{enumerate}
\item
the choice of the mathematical form of the MIM
\item
the choice of experimental data used to determine the parameters of the MIM, and hence the FF
\item
lack of the availability and appropriateness of a simulation approach to calculate certain solution properties
\end{enumerate}
For completeness, we should also include the general problem of MIM species cross-interaction parameters for mixtures. A large number of empirical combining rules has been proposed, but their suitability is unknown {\it a priori} and usually requires fitting to at least one experimental point to make the FF work reasonably well. Finally, some approaches focus on a limited goal of a MIM.  An extreme example of this approach is expressed by the following comment from a member of the biocomputing community \cite{vanGusteren2014}: ``$\dots$ {\it model can be kept simple, because it only needs to represent water at physiological thermodynamic conditions, $\dots$ and not all properties of a water model are equally important when serving as biomolecular solvent}". Such targeted models are not intended to contribute to improved general understanding of the behavior of aqueous solutions and are not considered further herein.\\

Concerning factor (1), any MIM includes, regardless of the considered system, van der Waals (non-electrostatic) and electrostatic interactions, denoted by ``vdW" and ``elect", respectively. A large number of possible combinations of MIM choices has led to ``{\it a large zoo of the ion parameters}" \cite{Horinek2009}. As a result, it is rather difficult to assess the performance of individual FFs and draw general conclusions on their accuracy and suitability.  Another important factor contributing to this zoo is that the vast majority of FFs have been developed on the assumption of pair--wise additive MIMs (further discussed in  Section \ref{section2}. Since this assumption is an approximation which is intrinsically incorrect in the case of mixtures, the developed FFs cannot perform satisfactorily, and this results in continuous attempts to improve them. However, sufficient evidence has been gathered showing that further progress can be achieved only by an explicit inclusion of many--body effects. \\

Concerning factor (2), once a MIM has been selected, its parameters must be determined by fitting to a set of experimental property data (referred to henceforth as the Training Set, TrS) to determine the FF.  The resulting FF is then tested by its prediction of data other than that contained in the TrS (referred to henceforth as the Test Set, TeS).  The choice of TrS  is a delicate problem, since the properties must be:
(1) sensitive to the MIM parameters,
(2) available experimentally or from ab--initio quantum chemical calculations, and
(3) computationally available by means of a molecular simulation algorithm. Our observation is that, in many cases, researchers have not viewed their FF parameter determination methodologies in this systematic way, and it seems that the choice of TrS is often governed by the availability and convenience of the features of available software packages.\\

Concerning factor (3), although a large number of both commercial and freeware simulation packages are available, they may not include methodologies to readily calculate certain properties. For example, recent aqueous electrolyte solubility studies have produced widely different results, even when using the same water and electrolyte FFs; this is illustrated in Table \ref{solubilitydata} in Section \ref{section5}, which shows results obtained by several research groups for NaCl aqueous solubility at ambient conditions. \\

In this perspectives paper, we  assess recent activities concerning the suitability and usefulness of the most commonly used both pair--wise additive and non-additive FFs for aqueous solutions of the most typical solution, that of NaCl. Specifically, (1) we discuss the way the FFs were developed, (2) the TrSs employed, and (3) their predictions of chemical potentials over the entire solution concentration range at ambient conditions and of the solubility (either as a TrS or a TeS). The last is particularly important, since the concentration--dependent chemical potential is one of the most important solution properties, since its knowledge as a function of $T$ and $P$ allows in principle the calculation of all other thermodynamic properties.
We focus on simulation results for the electrolyte chemical potential, solubility and activity coefficients, and for the water chemical potential and properties related to it. In the process, we discuss the correct evaluation of uncertainties for properties considered that arise from the combination of uncertainties of more than one simulation quantity.  We also propose a new method to directly calculate the standard  chemical potential in the Henry--Law ideality model by means of computer simulation.
We identify weaknesses and limitations of the FFs and outline our view of potential future developments with respect to the above factors.
\\

The next section of the paper provides a general description of the current and potential MIMs and FFs for water and for aqueous electrolytes, respectively.  The subsequent section summarizes and discusses the methodology used for using experimental data with which to fit the FF parameters.  The following section discusses recent calculations of various aqueous electrolyte
solution
properties, and the final section summarizes the current state of affairs and suggests future directions.


\section{Molecular Interaction Models and Force Fields}\label{section2}

Recent advances in computer technology have enabled researchers to consider more and more complex and realistic MIMs using molecular simulation methodology. Furthermore, from the end of the last century, developments in quantum chemistry and computing facilities have made it possible to model molecules at increasingly finer resolutions and to view them as geometrical objects, either rigid or flexible, with interaction sites embedded within  (usually, but not always) the atoms comprising the molecules and which can interact with interaction sites on other molecules. These interactions include, in general, electrically neutral van der Waals  forces, comprising both strong short ranged repulsions and medium ranged attractions (dispersion forces), long-ranged electrostatic interactions between permanent charges distributed within the molecules, and induced electrostatic interactions.  For flexible molecular models, intramolecular interactions must also be included, but this is rarely used for water or aqueous electrolytes (but see, {\it e.g.}, \cite{Gonzales2011} and references therein).\\

A general form of the interaction energy between two rigid molecules or ions $k$ and $\ell$ in a system of $N$ particles is thus
\begin{equation}
u({\bf q}^{(k)},{\bf q}^{(\ell)}) = \sum_{i,j}u_{\vdW}({\bf r}^{(k)}_i,{\bf r}^{(\ell)}_j) +
 \sum_{i,j}u_{\rm elect}({\bf r}^{(k)}_i,{\bf r}^{(\ell)}_j)
\label{ugen}
\end{equation}
where ${\bf q}^{(k)}$ represents a set of generalized coordinates of molecule $k$, $\vr^{(k)}_i$ is the position vector of site $i$ of molecule $k$, and the summations run over all site--site pairs $i$--$j$ of sites $i$ on molecule $k$ and sites $j$ on molecule ${\ell}$.
The total configurational energy, {$\cal{U}$} is given by
\begin{equation}
{\cal {U}} = \sum_{k,\ell} u({\bf q}^{(k)},{\bf q}^{(\ell)}) + u_{\rm nonadditive}({\bf q}^{(1)},{\bf q}^{(2)},\ldots,{\bf q}^{(N)}) \label{ugenfinal}
\end{equation}
The last term in Equation (\ref{ugenfinal}) corresponds to the induced electrostatic energy, which depends on the configuration of all the system particles..
MIMs differ in their different choices for the individual terms. Moreover, every MIM must also specify the molecular geometry, which further increases the number of possible site--site interactions.  The majority of efforts to develop FFs have been motivated by simplicity and, as a result, in the underlying MIMs the last term in Equation (\ref{ugenfinal}) has been neglected, assuming that non--additive effects may be incorporated implicitly by adjustment of the pairwise interaction parameters to suitable experimental data. Such FFs thus assume pairwise additivity of the interaction terms. \\

The most common choice for the vdW term in Equation (\ref{ugen}) has been the Lennard-Jones (LJ) MIM, which may
be
expressed in various equivalent forms:
\begin{eqnarray}
u_{\vdW} \equiv u_{\LJ}(r)  & = & 4\epsilon
  \left[\left(\frac{\sigma}{r}\right)^{12} -
        \left(\frac{\sigma}{r}\right)^6\right]  \label{uLJ}
\\
& = & \epsilon
  \left[\left(\frac{r_m}{r}\right)^{12} -
        2\left(\frac{r_m}{r}\right)^6\right]\\
&=& \frac{C_{12}}{r^{12}} - \frac{C_6}{r^6}
\end{eqnarray}
where $\sigma$ and $\epsilon$ are energy and size parameters, respectively, and $r_m$ is the separation distance at the minimum.
This provides a reasonable approximation at ambient conditions, but is known to be unsuitable at higher $T$ and $P$, where the real repulsive interactions are much softer.
This plays an important role in electrolytes, for which the strong electrostatic attraction between unlike ions results in shorter distances between particles. An alternative choice to the LJ model with a sounder physical basis is the Buckingham (EXP6) MIM:
\begin{eqnarray}
u_{{\rm EXP6}}(r) & = & \frac{\epsilon}{1-6/\alpha_{\rm EXP6}}
  \left[\frac{6}{\alpha_{\rm EXP6}}\exp^{\alpha_{\rm EXP6}(1-r/r_m)} - \left(\frac{r_m}{r}\right)^6\right]
 \\
& = & A\exp^{-Br} - \frac{C_6}{r^6} \label{uexp6}
\end{eqnarray}
In addition to the two adjustable parameters ($\epsilon$ and $r_m$)  scaling the interaction energy and distance, respectively, it employs a third parameter, $\alpha_{\rm EXP6}$, which determines the softness of the short-range repulsion.  A simple extension adding an additional attractive term \cite{Ferrario2002,Aragones2012,Sanz2007} has also been used for electrolytes (Born--Mayer--Huggins--Tosi--Fumi MIM; BMHTF)\cite{Fumi1964,Tosi1964,Huggins1933,Mayer1933},
\begin{equation}
u_{{\rm BMHTF}}(r) = u_{{\rm EXP6}}(r) - \frac{C_8}{r^8}
\label{uBMHTF}
\end{equation}
Other possible choice is the Mie $(n-m)$ MIM~\cite{Mie1903} which generalizes the exponents (6,12) in the LJ MIM  to $(n,m)$.  However, this possibility has not been thus far considered for electrolytes.\\

Early models  for the electrostatic interaction in Eqn. (\ref{ugen}) considered point charges at molecular sites:
\begin{equation}
u_{\Coul}(r) = \frac{1}{4\pi\epsilon_0}
  \frac{q_k q_l}{r}
\label{uCoul}
\end{equation}
where $q_k$ is the partial charge of site $k$ and $r$ is the separation distance between sites.  However, charges within a molecule cannot be strictly localized, and more plausible models consider a smeared charge ({\it e.g.}, a Gaussian distribution) localized on the respective sites \cite{ChialvoCummings1998}.
Additionally, to better estimate the electric field created by molecules, it is also possible to add to the point charge higher permanent multipoles as in, {\it e.g.}, the AMOEBA model for water \cite{RenPonder2004}. \\

From the physical point of view it is evident that FFs based on the assumption of pairwise additivity (neglecting the last term in Equation (\ref{ugenfinal})) are inadequate in principle. Such a FF is not able to account simultaneously for the predominantly two--body interactions in the dilute gas phase and the many--body interactions at liquid-like densities without invoking state--dependent FF parameters. Such FFs are similarly deficient for mixtures, since the interactions between molecules of a given type are significantly affected by the presence of molecules of different types.
It is easy to appreciate that in the dense phase of a polar substance such as water the electronic structure of its molecules must become significantly distorted (polarized) by the effect of the surrounding molecules. However, even in a gas--phase environment, polarizability can strongly affect the intermolecular energy \cite{Caldwell1955}. \\

Early attempts to go beyond pairwise additivity date to the 1970s \cite{Vesely1977,Stillinger1978,Barnes1979,Warshel1979} and it has become increasingly recognized that non--additive many--body effects must be incorporated into MIMs to bring them closer to reality, see, e.g.,  \cite{DyerCummings2008,SalaGuardia2010}. Among the possible non--additive contributions, the theoretical treatment of short--range three--body repulsion is much less certain than three-body dispersion and has thus been omitted in all models. In contrast, it has been well documented that multibody dispersion interactions can be adequately described using the Axilrod-Teller triple dipole term \cite{LiSadus2007}, but nonetheless, this type of non--additivity has been shown to have only a small effect on the properties of systems involving water and is negligible in comparison with the effects of the induced long-ranged electrostatic interaction, {\it i.e.}, polarizability \cite{LiSadus2007}. Polarizability has thus been the only term contributing $u_{\ind}$ to account explicitly for the distortion of the electronic structure of molecules exposed to an external field.\\

Polarizability can be considered in a number of forms.  Early models used a polarizable point dipole or a fluctuating charge \cite{Rick1994}. The induced dipole of a molecule, $\bf p$, is proportional to the total electric field at its location due to all other molecules and is given by
\begin{equation}
{\bf p} = \alpha_{\rm pol} ({\bf E}^q + {\bf E}^p)
\end{equation}
where $\alpha_{\rm pol}$ is molecular polarizability, ${\bf E}^q$ is
the electric field due to the permanent charges or dipoles on other molecules of the system and ${\bf E}^p$ is the electric field due to the induced dipoles of other molecules. \\

Another possibility is to use a charge-on-spring MIM. In this case, the charge may be either in the form of a point charge, $q_{\rm D}$, (the so--called Drude particle, which is a massless virtual site \cite{Drude1902}) or a Gaussian distribution. In the former case, the induced dipole is ${\bf p}=\alpha_{\rm pol}{\bf E}=q_{\rm D}{\bf l}$, where $\alpha_{\rm pol}$ is the polarizability, $\bf l$ is the elongation vector of the spring and $\bf E$ is the total field acting on the charge $q_{\rm D}$. The polarization contribution of molecule $i$ to the total energy is then
\begin{equation}
u_{\qSpring} =  \frac{1}{2} k_{\rm D}{\bf l}_i^2
\label{qSpring}
\end{equation}
where the spring constant is given by $k_{\rm D}=q_{\rm D}^2/\alpha_{\rm pol}$, and similarly for the Gaussian distribution. \\

Vega and Abascal~\cite{Vega2011} discuss at length the limitations of FFs based on rigid non-polarizable water MIMs with respect to a number of properties.
Although the available models could be, at least in principle, improved, {\it e.g.}, by replacing the traditional LJ MIM by the more realistic EXP6 MIM,
and/or the point charge by a Gaussian distribution,
\begin{equation}
\rho_{i}(\vr)=q_{i} \exp \left[-\alpha_{i}^2 \left(\vr - \vr_{i}\right)^2\right]   \,\,,
\label{qGauss}
\end{equation}
they conclude that future development towards better water MIMs should proceed via the inclusion of polarizability.
In Equation (\ref{qGauss}), $\rho_{i}$ is the charge density associated with site $i$, $\vr_{i}$ is the position of the site, and the interactions are given by the electrostatic potential energy of two Gaussians,
\begin{equation}
u_{{\rm GC}, ij} =\frac{1}{4\pi\varepsilon_0 } \sum_{a,b} \frac{q_{ia}q_{jb} }{|\vr_{ia}-\vr_{jb}|}{\rm erf} \left(\alpha_{iajb} |\vr_{ia}-\vr_{jb}| \right)
  \label{uGauss}
\end{equation}
where
\begin{equation}
\alpha_{iajb}= \frac{\alpha_{ia}\alpha_{jb} }{\sqrt{ \alpha_{ia}^2 + \alpha_{jb}^2}}
\end{equation}
We mention in passing that changing the geometry of the water MIM or adding additional sites may be other alternatives to improve the performance of pairwise FFs; see, {\it e.g}., \cite{YuRoux2013}. \\

Within the above described MIMs and in particularly the case of mixtures of different molecules, it is possible prescribe formulae relating the interaction terms involving unlike sites to those involving the underlying like sites, by means of so--called combining rules.  For a mixture, one purpose of combining rules is to decrease the number of parameters to be determined to define the FF, which applies to the case of aqueous electrolytes, where even in the case of a single electrolyte are ternary mixtures of water, cations and anions. One must also approach the problem cautiously for such solutions, in which the interactions between unlike sites, {\it i.e.}, the ion--water interactions and the unlike--ion--ion interactions play a crucial concentration--dependent role.  For example, some ion FF parameters were developed using TrS properties at infinite dilution, where the ion--water interactions predominate, and the ion--ion interactions were subsequently obtained by means of prescribed combining rules ({\it  e.g.}, the approaches of Horinek \ea \cite{Horinek2009} or of Lamoureux and Roux \cite{AHSWM4DP2006}).  This methodology is unlikely to yield an accurate description of the solution behaviour at higher concentrations and of the crystalline solid, where ion--ion interactions are more important.\\

Combining rules are MIM specific, and the overwhelming majority has been developed for the LJ MIM. The most common  is the Lorentz rule for the diameter and the Berthelot rule for the energy (referred to as the L--B rules),
\begin{eqnarray}
\sigma_{ij} &=& \frac{1}{2}(\sigma_{ii} + \sigma_{jj})  \\
\epsilon_{ij} &=& \sqrt{\epsilon_{ii}\epsilon_{jj}}
\label{LBrule}
\end{eqnarray}
However, the Lorentz rule (arithmetic mean) has sometimes been replaced by the Berthelot rule (geometric mean) for $\sigma_{ij}$ and the geometric mean combining rule has thus been used for both energy and size cross parameters:
\begin{equation}
\sigma_{ij} = \sqrt{\sigma_{ii}\sigma_{jj}}
\end{equation}
The above rules are known not to perform well, and it has become a general practice to empirically modify the $\epsilon_{ij}$ parameter,
\begin{equation}
\epsilon_{ij} = (1-c_{ij})\sqrt{\epsilon_{ii}\epsilon_{jj}}
\end{equation}
where $c_{ij}$ is usually (but not always) close to zero. The approach of fitting $c_{ij}$ to one experimental data point of the mixture usually improves the result considerably. However, since it is well--known that fluid properties are very sensitive to the excluded volume (dependent on $\sigma$), see, e.g. \cite{Rouha2009}, an empirical adjustment to $\sigma_{ij}$ might rather be more appropriate.  Another possibility is the combining rule of  Kong~\cite{Kong1973a}, which expresses the repulsive interaction between unlike atoms as the sum of deformation energies. Its extension to the full EXP6 is the Kong-Chakrabarty rule \cite{Kong1973b}.


\subsection{Water}\label{section2.1}

In view of water's importance, many MIMs on which water FFs are based have been proposed at various levels of molecular--level realism. As previously mentioned, in addition to the many possible choices for the individual site-site interactions one must also consider the geometry of the water molecule. With only a few exceptions (see, e.g., \cite{NadaEerden2003}), contemporary models consider the following geometries: three sites mimicking positions of the central oxygen atom and two hydrogens (e.g., TIP3P \cite{SPCTIP3P}, SPC \cite{SPCTIP3P}, or SPC/E \cite{SPCE1987}); four sites with an additional site (referred to as an $M$-site) located on the H--O--H bisector and bearing a negative charge (the family of TIP4P models \cite{TIP4P1983}); five sites, with two additional negatively charged sites representing a lone electron pair (TIP5P models \cite{TIP5P}). Also, most common water models do not consider the hydrogen sites as ones generating also non--Coulombic interactions, bearing only charge. The development of non-polarizable water models seems to have culminated with the development of a series of TIP4P--type models \cite{TIP4P2005,TIP4PEw} with TIP4P/2005 emerging as the overall champion when considering a range of TeS properties~\cite{Vega2011}.  \\

\begin{figure} 
 \centering
\includegraphics[scale=1.0]{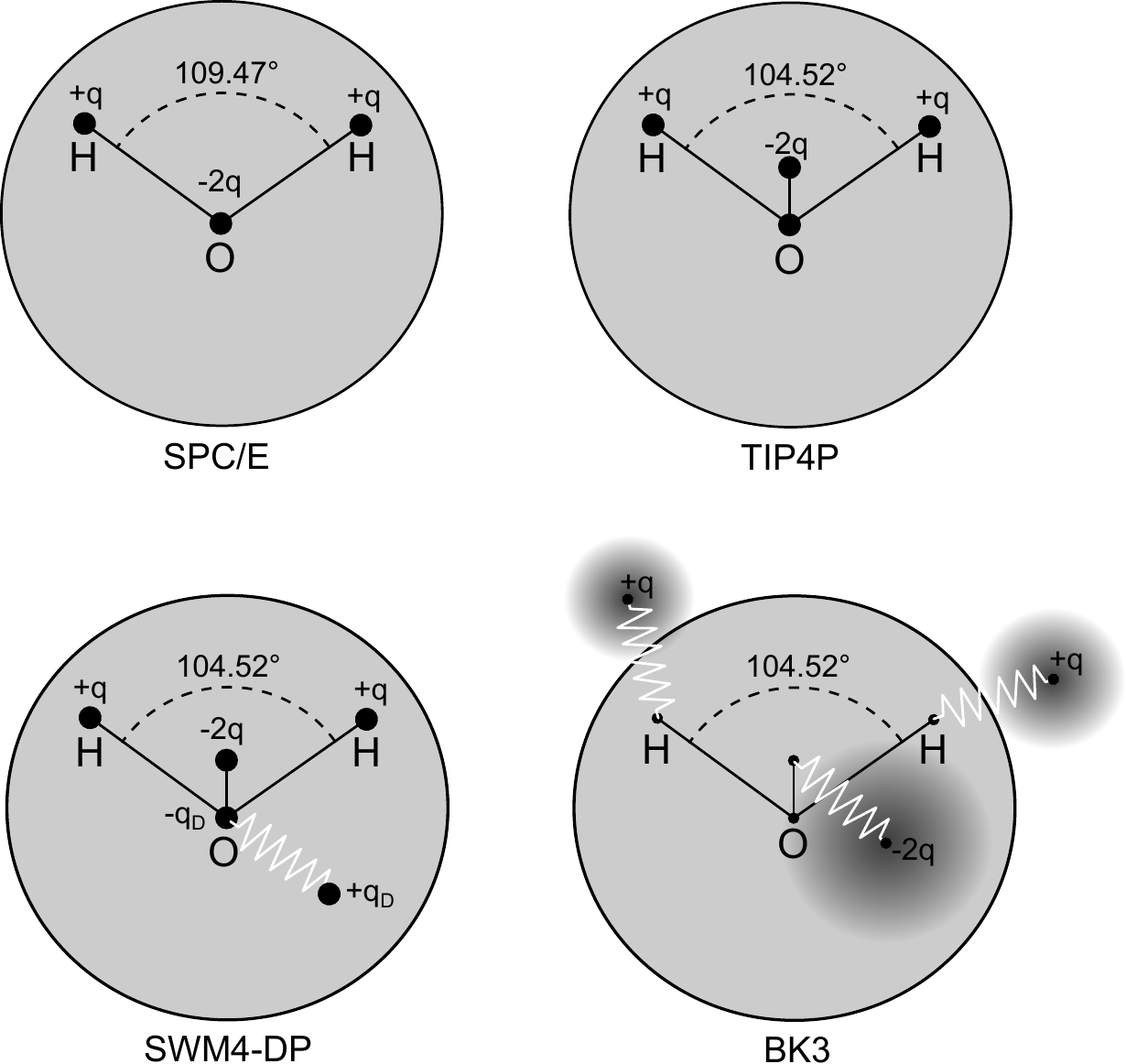}\vspace*{3cm}
 \caption{Geometry of the nonpolarizable water three-site SPC/E \cite{SPCE1987} and four-site TIP4P \cite{TIP4P1983} molecular interaction models, and the polarizable SWM4-DP \cite{SWM4DP2003} and BK3 \cite{BK32013} FFs. The rigid backbone of each FF is denoted by the connecting lines, the grey circles correspond to the van der Waals interaction, the large black dots represent point charges, the blurred circles Gaussian charges, the white zigzag lines are harmonic springs acting between the massless Drude charges and the rigid backbone.}
 \label{fig:geometry-water}
\end{figure}

The MIMs of the most commonly used models for water, the SPC/E and TIP4P models, are shown in the first row of Figure \ref{fig:geometry-water}, along with some of the FF parameters. For the remaining parameter values of the FFs, we refer the reader to the original papers. It may be surprising that all these models employ the traditional LJ potential for the non-electrostatic interactions, in spite of its known limitations. The likely reason for this choice is that it is much less computationally intensive in comparison to the EXP6 model repulsion, which was an important factor several decades ago. \\

The list of polarizable water FFs is relatively long, but only a handful of them have also been used for solutions of electrolytes. Models developed by the end of the last century were reviewed by Halgen and Damm \cite{Halgen2001}. With only a few exceptions (see, e.g., \cite{Bernardo}), these were simply polarizable versions based on available non--polarizable FFs obtained by adding a polarizable point dipole. A different path towards qualitatively different MIMs began with the GCPM model of Chialvo and Cummings \cite{ChialvoCummings1998}, in which the vdW interaction is replaced by the EXP6 potential and the point charges are replaced by a Gaussian distribution. This original model was then improved by the reparametrization by Paricaud et al. \cite{Paricaud2005}.\\

Other FFs falling into this category, {\it i.e.}, based on MIMs going beyond the traditional choice of the LJ potential and point charges, and used also for electrolyte solutions are the recent FFs of Lamoureux and Roux (SWM4-DP) \cite{SWM4DP2003} and of Baranyai and Kiss (BK3) \cite{BK32013}. SWM4--DP maintains the LJ model for the vdW interaction and fixed charges at sites, but the polarizability is modeled by a charge-on-spring model. The latest model which goes systematically beyond the LJ potential and point charges is the BK3 FF. Moreover, because the permanent charges of the sites are incorporated into the charge-on-spring it also reduces the total number of parameters to be adjusted to experimental data.
The geometries of the SWM4-DP and BK3 force fields are shown in the second row of Figure \ref{fig:geometry-water}, along with some of the FF parameters, the remaining values of which are given in the original papers.
\subsection{Simple electrolytes}\label{section2.2}

There is a plethora of electrolytes including large complex ions comprised of many different atoms, however, the current state of understanding their microscopic behavior limits reasonable applications of their force fields only to electrolytes of the simplest monovalent alkali metal and halide ions. Even these electrolytes pose a great challenge for the quantitative prediction of thermodynamic properties. We hence restrict our consideration here primarily to such systems.\\

Considering the physical nature of ions, their MIMs must reasonably incorporate the following features: (1) the van der Waals interaction of the ion with particles of solvent and with other ions, (2) the electrostatic interaction of the total charge of the ion with other particles, (3) the polarizability of the ion, and (4) combining rules that can be in principle omitted for the price of introducing additional free parameters of interactions between unlike sites.\\

The interactions in (1) and (2) can be modeled similarly to approaches described in the preceding section concerning
water FFs. Most ionic FFs employ the LJ MIM for (1), Equation~(\ref{uLJ}), and for (2) a point charge placed to the center of the ion with its magnitude equal to the total charge of the ion, interacting by the Coulomb law, Equation~(\ref{uCoul}).  Exceptions are the FFs of Kiss and Baranyai \cite{AHBK32014}, which use the Buckingham MIM (Equation~(\ref{uexp6}) for (1) and Gaussian charges, Equation~(\ref{uGauss})  for (2)). An earlier exception  used by several authors in their pioneering studies of solubility \cite{Ferrario2002,Sanz2007,OEMC2011,Paluch2010,Paluch2012} models the van der Waals interactions between water and ions by the LJ MIM but the ion--ion interactions by the Born--Mayer--Huggins--Tosi--Fumi MIM (Equation (\ref{uBMHTF})).\\

\begin{figure} 
 \centering
\includegraphics[scale=1.0]{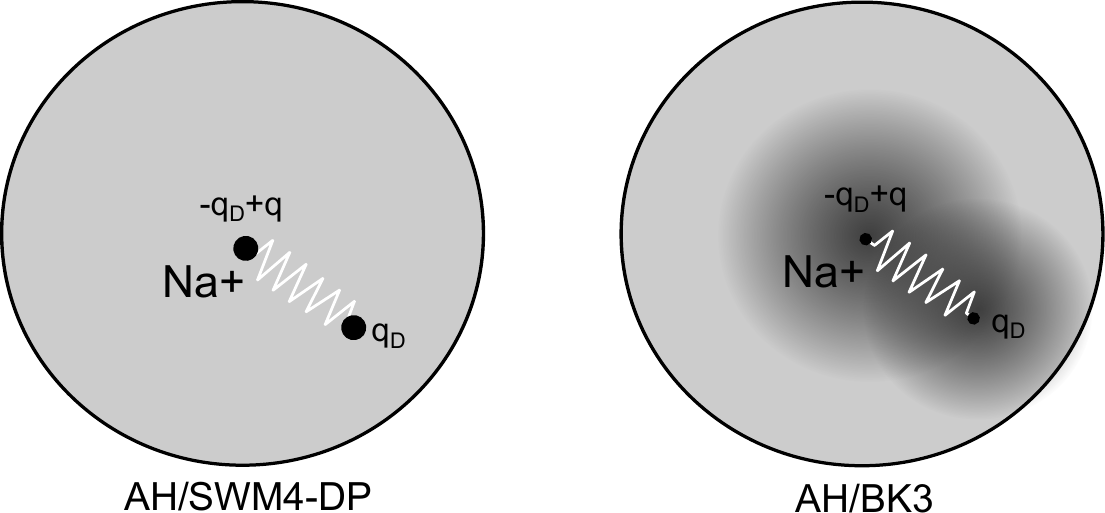}\vspace*{3cm}
 \caption{Geometry of the polarizable AH/SWM4-DP~\cite{AHSWM4DP2006} and AH/BK3~\cite{AHBK32014} molecular interaction models for Na$^+$. The grey circles correspond to the van der Waals interaction, the large black dots represent point charges, the blurred circles Gaussian charges, the white zigzag lines harmonic springs acting between the massless Drude charges and the centers of the ions.}
 \label{fig:geometry-ions}
\end{figure}

Regarding (3), the polarizable ionic FFs  of  Lamoureux \ea \cite{AHSWM4DP2006,Yu2010} and of Kiss and Baranyai \cite{AHBK32014} model polarizability by the charge-on-spring approach.  Those of Lamoureux \ea model the electrostatics of each ion by two point charges of magnitudes $q_{\rm D}$ and $q - q_{\rm D}$ fixed at the center of the ion and the other charge $q_{\rm D}$ attached to the center by a harmonic spring, with two variants AH/SWM4--DP \cite{AHSWM4DP2006} and AH/SWM4--NDP \cite{Yu2010} differing in the sign of their $q_{\rm D}$. A similar approach using a positive $q_{\rm D}$ value was used in the AH/BK3 FF of Kiss and Baranyai \cite{AHBK32014}, but employed Gaussian charges. The geometries of the AH/SWM4--DP and AH/BK3 force fields for Na$^+$ are shown in Figure \ref{fig:geometry-ions}, along with some of the FF parameters, the remaining values of which are given in the original papers.\\

\section{Electrolyte FFs}\label{section3}
\subsection{FF Training sets}\label{section3.1} 

Since the values of all thermodynamic functions from the set $\{P,V,T,U,S,H,G,A\}$, where $U$ is the total internal energy, $S$ is entropy, $H$ is enthalpy, $G$ is the Gibbs function, and $A$ is the Helmholtz free energy, are fixed by fixing any two property values, one choice of TrS would be 2 of them, ideally in terms of their fundamental variables, {\it i.e.}, $U(S,V),H(S,P),G(T,P),A(T,V)$, the most convenient choice of which would be one of the final two. If a simulation approach could match the experimental values of these as a function of the system composition, then all simulated properties would match the experimental results for all thermodynamic properties.   Since this is a difficult task, the TrS could be augmented by values of the property derivatives.  We note that any first--order derivative involving the eight thermodynamic functions listed can be expressed in terms of three of their independent first derivatives \cite{Bridgman1914}; for example, $(C_p,C_v, \chi_T)$ might be chosen.\\

The properties employed in the TrS
and the TeS
for aqueous electrolyte FFs have typically been selected from the following list:
\begin{enumerate}
\item density as a function of concentration 
\item individual ion and total electrolyte hydration free energy at infinite dilution 
\item individual ion and total electrolyte hydration entropy at infinite dilution 
\item gas phase single ion--water and cluster properties 
\item crystal density 
\item crystal lattice energy 
\item crystal chemical potential 
\item chemical potential derivative and partial molar volumes at a moderate concentration (related to  Kirkwood--Buff integrals; see \cite{Weerasinghe2003,Smith2010,Gee2011}) 
\item liquid structural properties (coordination numbers; pair correlation function)
\item chemical potential as a function of concentration 
\item electrolyte solubility 
\item osmotic pressure, $\Pi$, as a function of concentration 
\item
enthalpy of mixing as a function of concentration 
\item
transport properties (diffusion coefficients, viscosity, conductivity, permittivity, dielectric constant)
\end{enumerate}

It is important that properties within the TrS be sensitive to the FF parameter values, and that they reasonably cover the entire experimentally accessible concentration range.  For example, the density (Property 1) is primarily dependent on the molecular sizes rather than on the energy parameters, and it is important that some higher concentration property also be included.   However, although crystal properties (Properties 5--7) have sometimes been considered, the majority of researchers have used infinite dilution properties (Properties 2--4), in part because these are readily calculated in widely available computer packages.   Only Properties 8--12 relate to higher solution concentrations.\\

Concerning Property (2), the use of experimental single--ion hydration free energy data as a TrS property merits more detailed discussion, in view of recent work of the Chialvo group \cite{Vlcek2013}.  They argue that such data may be unsuitable because their values are not directly experimentally accessible, but are obtained indirectly based on different experiments in conjunction with some extra--thermodynamic assumptions whose validity is questionable. One of the most reliable sources of single--ion hydration properties typically considered are data that rely on the cluster--pair--based approximation technique of Tissandier~\ea \cite{Tissandier1998}. The validity of this approach was recently studied by Vl\v{c}ek \ea \cite{Vlcek2013} by a set of molecular simulations of single--ion hydrated clusters of different sizes and an infinite flat water-vapor interface. They found that the hydration properties of single cations and anions in growing clusters converge to their limits rather slowly, which is incompatible with the cluster--pair--based approximation. They thus conclude that there can be significant errors in the experimental single--ion bulk solution data. In a subsequent study, Vl\v{c}ek \ea \cite{Vlcek2015} used several water FFs to evaluate their accuracy in predicting hydration thermodynamics and structural properties of small clusters. They found that: (1) all the studied FFs provide comparable predictions for pure water clusters and cation hydration properties, but they differ significantly in their description of anion hydration properties; (2) none of the investigated classical FFs can quantitatively reproduce the experimental gas--phase hydration thermodynamics; and (3) a single polarizable site located at, or near, the oxygen center may not be sufficient to reproduce the experimental hydration thermodynamics.\\

Concerning Property (8), Kirkwood--Buff integrals (KBI) are infinite--ranged integrals of the pair correlation functions, $g_{ij}(r)$, in terms of which the concentration derivative of the salt chemical potential and the solution partial molar volumes may be expressed~\cite{BenNaim1992}.  A potential difficulty with using the KBIs as TrS properties is that the upper limit of the integrals must be truncated, requiring great care in their numerical calculation, although this drawback has recently been removed in principle by
Kr\"{u}ger~\ea\cite{Kruger2013}. It should also be noted that agreement with the derivative of a function does not guarantee that the function values themselves agree; hence if Property 8 is used it should be augmented with additional TrS properties.

\subsection{Recent electrolyte FF development}\label{section3.2}

Recent studies have used different combinations of the TrS properties of the previous section, sometimes accompanied by predictions of  properties in a TeS.  The following is a brief survey of recent studies:
\begin{enumerate}
\item 
Deublein \ea \cite{Deublein2012,Walter2012} considered the alkali metal ions (Li$^+$, Na$^+$, K$^+$, Rb$^+$, Cs$^+$) and the halide ions (F$^-$, Cl$^-$, Br$^-$, I$^-$) at $T=293.15$ K and $P= 1$ bar using a LJ vdW and point charge MIM in conjunction with the L--B combining rule and a range of water FFs. Their TrS was item (1) in the above list, in terms of the dependence of the reduced density $\tilde{\rho}$ (defined as the ratio of the solution density to the pure water density) on the salt mass fraction $w$, which is related to the molality by
\begin{equation}
w = \frac{mM_s}{1+mM_s}
\end{equation}
where $M_s$ is the molecular weight of the solute in kg mol$^{-1}$.  They used a linear approximation to $\tilde{\rho}$
\begin{equation}
\tilde{\rho} \equiv \frac{\rho}{\rho_w} = 1 + \alpha w
\end{equation}
where $\rho$ and $\rho_w$ are the mass densities of the solution and pure water, respectively, and $\alpha$ is an empirical constant. Finding that the solution reduced density was relatively insensitive to the ion LJ energy parameters, they fixed them at $\epsilon/\kB=100$ K~\cite{Deublein2012} or 200 K~ \cite{Walter2012}, where $\kB$ is Boltzmann's constant, and fitted the ion diameters.  They achieved good agreement with the TrS data; in the case of  NaCl, they considered $w \lessapprox 0.15$, corresponding to $m \lessapprox 3.0$.   \\

They considered a TeS consisting of various structural quantities (first maximum and first minimum of the ion--oxygen correlation function and coordination number), and found that the agreement with experiment was reasonable and that the ion parameters were transferable to several different water FFs in their ability to predict the reduced density.  They also considered a TeS of ion self--diffusion coefficients in water for several water FFs, but found the agreement with experiment to be relatively poor \cite{Walter2012}.  Later~\cite{Reiser2014}, they refitted the LJ energy parameter based on experimental self-diffusion coefficient data of the alkali metal cations and the halide anions in aqueous solution as well as the position of the first maximum of the radial distribution function of water around the ions. They used the electric conductivity, the hydration dynamics of water molecules around the
ions, and the enthalpy of hydration in their TrS.  The conductivity was predicted for aqueous NaCl and CsCl very well, and the predicted enthalpies of hydration for all alkali halides deviated from the experimental values between 10\%-20\% .\\

\item 
Joung and Cheatham~\cite{Joung2008,Joung2009} considered the alkali metal ions (Li$^+$, Na$^+$, K$^+$, Rb$^+$, Cs$^+$) and the halide ions (F$^-$, Cl$^-$, Br$^-$, I$^-$) at 298 K and 1 atm. in conjunction with a LJ plus point charge MIM and L--B combining rule and several water models (TIP3P \cite{SPCTIP3P}, TIP4PEw \cite{TIP4PEw} and SPC/E \cite{SPCE1987}). They fitted the ion $\epsilon$ and $\sigma$ values using a TrS consisting of items (2) and (6).  They also used a subsidiary TrS of item (5).  They  considered a TeS of structural quantities \cite{Joung2008}, and subsequently \cite{Joung2009} enlarged this to consider activity coefficients, diffusion coefficients, residence times of atomic pairs, association constants, and solubility. \\

They found that: (1) the reliability of the ion force fields is significantly affected by the specific choice of water model ({\it i.e.}, the electrolyte FFs are not transferable to different water models); (2) the ion-ion interactions are very important to accurately simulate the properties, especially solubility; (3) the electrolyte FFs compatible with SPC/E and TIP4PEw water are preferred for simulation in high salt environments compared to the other ion FFs.\\

\item 
Smith \ea~\cite{Weerasinghe2003,Smith2010,Gee2011} considered aqueous NaCl solutions at 300 K and 1~atm in conjunction with a LJ point charge MIM and either L--B or geometric mean combining rules augmented by a modified energy parameter for the Na$^+$--water oxygen interaction. They used item (8) at a solution concentration of 4 molar (4M) in the TrS, supplemented by values of the ionic radii consistent with crystal lattice dimensions (item (6)), the position of first peak of the ion--O distance in 1M and 4M solutions (item (9)). Fitting the KBIs enabled agreement of the FF with the density, partial molar volumes, and concentration derivative of the salt activity coefficient, and they considered a TeS of relative permittivity, enthalpy of mixing and diffusion constants, finding reasonable agreement with experiment.  By incorporating a known empirical expression for the experimental salt activity coefficient at low concentrations into their analysis of their results, they achieved good agreement with experiment of the activity coefficient at higher concentrations. \\

\item 
Horinek {\it et al.} \cite{Horinek2009} considered the alkali metal ions (Li$^+$, Na$^+$, K$^+$, Cs$^+$) and the halide ions (F$^-$, Cl$^-$, Br$^-$, I$^-$) at 300 K and 1 bar in conjunction with a LJ point charge electrolyte MIM and SPC/E water.  As TrS they selected items (2) and (3) for each ion relative to that of Cl$^-$, for which they used $\sigma_{\rm ion-oxygen}$ and $\varepsilon_{\rm ion-oxygen}$ of Dang \cite{SmithDang1994}.  They then selected three reasonable values of  the corresponding cation $\varepsilon$ parameters. They used item (9): the position of the first peak of the ion--O distance at infinite dilution as primary TeS. They found that the TrS was quite accurately reproduced, and the agreement with the TeS was reasonable.\\

\item 
Reif and H\"{u}nenberger \cite{Reif2011a} considered the alkali metal ions (Li$^+$, Na$^+$, K$^+$, Rb$^+$, Cs$^+$) and the halide ions (F$^-$, Cl$^-$, Br$^-$, I$^-$) at 298.15 K and 1 bar in conjunction with a LJ point charge electrolyte model and with SPC and SPC/E water. They fitted the ion--water FF parameters to the infinite dilution TrS properties of the ion--ion LJ $C_6$ parameters and item (2) (see \cite{Reif2011a} for details), developing three parameter sets corresponding to three different values of the proton hydration free energy.  They considered three different combining rules for the ion--ion parameters, and their TeS consisted of the infinite dilution electrolyte partial molar volumes and the first peak in the ion--water radial distribution function, in addition to the salt lattice parameters and energies.  Their parameter sets achieved reasonable qualitative agreement with these properties.\\

\item 
Fyta {\it et al.} \cite{Fyta2010, Fyta2012a,Fyta2012b} considered different combinations of TrS properties to determine FFs based on the LJ point charge MIM for alkali halides using the SPC/E H$_2$O FF. In \cite{Fyta2010}, they fixed the FF parameters for Na$^+$ and Cl$^-$ to those of Dang~\cite{SmithDang1994} and  used items (2) and (12) in conjunction with L--B rules. They found difficulties for FFs involving the I$^-$ and F$^-$ ions. In  \cite{Fyta2012a}, they used items (2) and (8) and relaxed both the size and energy L--B cation--anion combining rules.  In  \cite{Fyta2012b}, they used items (2) and (8) and further examined this approach for the problematic cases involving NaF and KF,  even suggesting concentration dependent variations.\\

\item 
Lamoureux and Roux \cite{AHSWM4DP2006} used their polarizable SWM4--DP water FF \cite{SWM4DP2003} as a basis to develop a set of polarizable FFs for the akali metal ions (Li$^+$, Na$^+$, K$^+$, Rb$^+$, Cs$^+$) and the halide anions (F$^-$, Cl$^-$, Br$^-$, I$^-$) at 298.15~K and 1 atm.  They used the MIM described in
in Section \ref{section2.1},
with L--B combining rules for the unlike interactions. Their TrS was item (2) and scaled gas--phase values of polarizability corresponding to ion--monohydrate binding energies and distances (see the paper for details). Their FFs achieved good agreement with this data. Their TeS was item (9). They found that the cluster energies agreed well with calculated quantum mechanical results and the structural results were acceptable. They later \cite{Yu2010} applied the same approach using  their SWM4--NDP water model, including additionally the alkaline earth cations (Mg$^{2+}$, Ca2$^{2+}$, Sr2$^{2+}$, Ba$^{2+}$, Zn$^{2+}$).\\

In subsequent work~\cite{Luo2010,Luo2013}, they added the osmotic pressure (item (12)) to the TrS and adjusted the cation--anion LJ parameter to fit the data at moderate densities, based on an MD algorithm they developed for $\Pi$ in \cite{Luo2010}.  The incorporation of $\Pi$ as a TrS property based on their approach was later taken up by Saxenea and Garcia~\cite{Saxena2015}, who used it to develop FFs for MgCl$_2$ and CaCl$_2$.\\

\item 
Kiss and Baranyai~\cite{AHBK32014} used their polarizable BK3 water FF~\cite{BK32013} as a basis to develop a set of polarizable FFs for the akali metal ions (Li$^+$, Na$^+$, K$^+$, Rb$^+$, Cs$^+$) and the halide anions (F$^-$, Cl$^-$, Br$^-$, I$^-$) at 298~K and 1 bar.  They used the MIM described in
Section \ref{section2.1}; they used the Kong combining rule~\cite{Kong1973a} for the repulsive portion parameters $A$ and $B$, and the Berthelot geometric rule for the $C_6$ parameter.
Their TrS was items (2) and (4), and their TeS was items (5), (6) and (9).    They found that, with the exception of the Li-halides, the densities and lattice energies of the salt crystals agreed with the experimental results to within 3\%.\\

\item 
Mou\v{c}ka \ea \cite{MNSLimitations2013} studied 13 FFs based on the LJ point charge MIM with L--B combining rules and tailored to SPC/E water FF \cite{SPCE1987} with respect to properties (1), (5), (10) and (11) in the previously listed set of potential TrS properties, and found that only those of  Joung and Cheatham \cite{Joung2008,Joung2009} and one of the FFs of Horinek {\it et al.} \cite{Horinek2009} produced reasonable, although not completely satisfactory, results. Among their findings was that many of the FFs exhibited salt  precipitation at concentrations well below the experimental solubility limit. In a separate  study \cite{MNSDevelopment2013}, they showed that electrolyte FFs based on this MIM  are unable to provide quantitatively accurate predictions of important higher concentration aqueous solution properties. They used Properties (1), (7), (10) and (11) to develop a new FF that
turned out to be
only marginally superior to all others based on this MIM.\\

\item 
Kim and coworkers~\cite{Kim2012} studied the effects of salt on the self-diffusion coefficient of water in different alkali halide aqueous solutions in the context of various FFs from the literature. They calculated the water self-diffusion coefficient as functions of concentration and temperature.  They considered five nonpolarizable water FFs with four nonpolariazable electrolyte FFs and two polarizable FFs, and different salts. They found that none of these FFs was able to reproduce the concentration dependence of the water self-diffusion coefficient for aqueous solutions of KCl, KBr, KI, CsCl, CsBr and CsI, whose experimental values are increasing functions of concentration; all simulation results for these solutions show a decreasing concentration dependence. They obtained only a slight improvement of this behaviour in the case of one of the polarizable FFs considered. In contrast, the qualitative dependence on temperature for both the water self--diffusion coefficient and the solution viscosity is reproduced very well by the simulations in comparison with experiment. They concluded that current MIMs are not acceptable for simple dynamic properties, and that more accurate MIMs should incorporate hydrogen-bonding explicitly or that the Lennard-Jones interaction should be replaced by a softer repulsion.\\

Later, Kann and Skinner~\cite{Kann2014} focused on this problem with the aim of improving the concentration dependence of the self--diffusion coefficient of water in aqueous alkali halide solutions. They modified the ionic charges by a scaling factor that ensures agreement with the experimental value of the dielectric constant of pure water.  They studied the typical examples of NaCl and CsI in detail, and combined this method with the ionic FFs of Mao and Pappu~\cite{Mao2012}, whose parameters are determined solely from the electrolyte crystalline solid.  They considered several non--polarizable water FFs, including the E3B FF~\cite{Leontyev2012}, which incorporates 3--body forces. They conclude that their charge scaling approach improves the concentration dependence of the water self--diffusion coefficient, particularly for NaCl.  They further studied density, ion--oxygen radial distribution functions and water and ion diffusion constants for a range of aqueous alkali halide FFs.  They concluded that the Mao and Pappu FF in conjunction with either the TIP4P/2005 or E3B water FF achieved the best results.
\end{enumerate}

\section{Aqueous electrolyte properties}\label{section4}

Since different research groups have obtained different results for identical problems, in the following we focus on results obtained using the same systems, thermodynamic conditions, FFs, and simulation protocols.  For example, in some of the earlier studies of Smith \ea~\cite{OEMC2005,OEMC2011,OEMC2012}, experimental values of the solid chemical potential~\cite{JANAF1998} were used
to determine solubilities
and in most cases the long--ranged electrostatic forces in the solutino phase were treated by means of the Generalized Reaction Field (GRF) approach~\cite{Hummer1996}.  Otherwise~\cite{OEMC2005,OEMC2011,OEMC2012,MNSLimitations2013,MNSDevelopment2013,MNSActivity2015}, they
conformed to the more common practices of employing the same FF for both the electrolyte and the solid, and the alternative (although not necessarily more scientifically sound; see, {\it e.g.}, \cite{Fukuda2012}) Ewald summation~\cite{AllenTildesley1987} approach for the electrostatics.
Except where noted otherwise, all comparisons to follow will refer to studies using this methodology.
The consequences of the different treatments of the electrostatics are discussed briefly in Section \ref{sub:solubility}.
In addition, many FFs exist in the literature for aqueous electrolytes, but only a very limited subset of them have been simultaneously studied by different authors for their predictions of chemical potentials and solubility.\\

We will thus primarily focus on results for aqueous NaCl at ambient conditions using the seven FFs listed below (listed in approximate chronological order), which have been studied by four different research groups: Vega \ea~\cite{Sanz2007,Aragones2012,Vega2015}, Paluch \ea \cite{Paluch2010,Paluch2012}, Panagiotopoulos \ea~\cite{Mester2015a,Mester2015b,Jiang2015} and W.R. Smith \ea~\cite{OEMC2005,OEMC2011,OEMC2012,MNSGibbsDuhem2013,MNSLimitations2013,MNSActivity2015}.  Five are  nonpolarizable FFs compatible with SPC/E water, based on the LJ MIM except as indicated, with an embedded point charge at the centre.  The remaining two are recently developed polarizable FFs.
\begin{enumerate}
\item
SD--SPC/E~\cite{SmithDang1994}:
This uses L--B combining rules for all LJ interactions.  The FF parameters are listed in \cite{OEMC2011}.
\item
KBI--SPC/E~\cite{Weerasinghe2003}:
This uses geometric mean combining rules for all LJ interactions except the Na$^+$--O interaction.  The FF parameters are listed in \cite{Weerasinghe2003}.
\item
SD--BMHTF--SPC/E: This uses a LJ MIM with embedded point charge at its center for the ion--water interactions with the Smith--Dang parameters~\cite{SmithDang1994} and the L--B combining rule, and for the ion--ion interactions the Born--Mayer--Huggins--Tosi--Fumi MIM of Equation (\ref{uBMHTF})~\cite{Fumi1964,Tosi1964,Huggins1933,Mayer1933}.  The FF parameters are listed in \cite{OEMC2011}.
\item
JC--SPC/E~\cite{Joung2008}:
This uses L--B combining rules for all LJ interactions.  The FF parameters are listed in \cite{Joung2008}.
\item
MNS--SPC/E~\cite{MNSDevelopment2013}:
This uses L--B combining rules for all LJ interactions; it was developed to be an improvement on the JC--SPC/E FF.  The FF parameters are listed in \cite{MNSDevelopment2013}.
\end{enumerate}

The polarizable FFs considered are the following:
\begin{enumerate}
\setcounter{enumi}{5}
\item
AH/SWM4--DP~\cite{AHSWM4DP2006}:
This polarizable FF is described in Section \ref{section2.2}.  The FF parameters are listed in \cite{AHSWM4DP2006}.
\item
AH/BK3~\cite{AHBK32014}:
This polarizable FF is described in Section \ref{section2.2}. The FF parameters are listed in \cite{AHBK32014}.
\end{enumerate}

\subsection{Solubility}\label{sub:solubility}

\begin{figure} 
\centering
\includegraphics[scale=0.9]{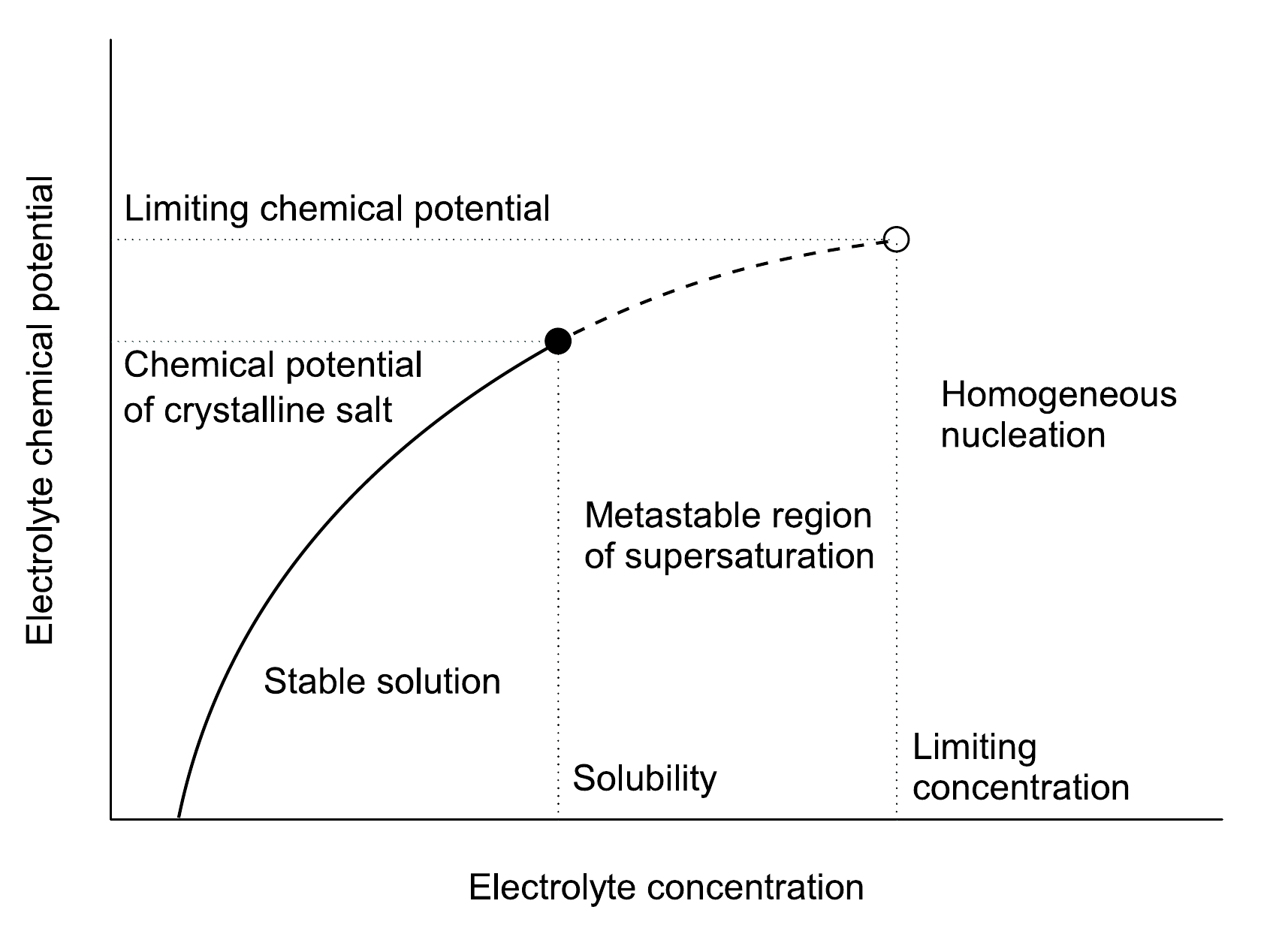}
\caption{Schematic representation of the relationship between solubility and electrolyte chemical potential.}
\label{fig:solubility-schematic}
\end{figure}

The electrolyte solubility is one of the most important aqueous electrolyte properties; we consider it first, since it provides a single quantity for each system and has been a particular focus of recent studies.  It is intimately related to the electrolyte chemical potential in the solution phase, $\mu_{\rm el}$, and in its pure crystalline state. The schematic in Figure \ref{fig:solubility-schematic} shows the general form of this relationship at a given system $(T,P)$. The solubility is the concentration at which the electrolyte chemical potential curve crosses the value corresponding to the solid.  Since the diffusional stability criterion
(see, {\it e.g.}, Equation (8.3.14) of \cite{OConnell2011})
states that at a given $(T,P)$ the chemical potential of a component in a binary mixture is an increasing function of its amount for a fixed amount of the other component, this means that
$\mu_{\rm el}$ is an increasing function of the solution molality, $m$.
At sufficiently low salt concentrations, we know that $\mu(m)$ is concave, and it seems reasonable that the qualitative behaviour in Figure \ref{fig:solubility-schematic} is generic for a strong electrolyte such as an alkali halide.  It shows $\mu_{\rm el}$ as a concave function, increasing from infinitely small values at low concentration, crossing the value of the crystal chemical potential, $\mu_s$, at a solubility limit $m_s$, then further increasing over a thermodynamically metastable concentration range until reaching a limiting chemical potential
at which homogeneous nucleation occurs, at a corresponding limiting concentration.  The ability of a FF to match the experimental solubility is a stringent test of its validity, involving the interplay of its ability to accurately predict both $\mu_s$ and the solution $\mu_{\rm el}(m)$ in its concentration vicinity.\\

In order for a FF to yield a realistic simulation involving an aqueous salt solution at a given $(T,P)$, its limiting concentration must lie above the concentration range of interest.  In order for the simulation to correspond to a thermodynamically stable solution, its solubility must also be greater than this range.  This is particularly important in biomolecular simulations, in which a salt solution forms the environment for the protein or other biomolecules of interest. If the solution concentration is beyond the limiting concentration of the FF, the solution is inherently unstable.  Use of a FF exhibiting a low solubility can result in pieces of crystal unknowingly being present in the simulation at physiological salt concentrations, a problem that has been alluded to in the literature
\cite{Auffinger2007,Joung2008}.\\

Algorithms for solubility calculations have been carried out by two different general approaches: (1) the thermodynamic approach (TA) of seeking the concentration $m_{\rm s}$ at which the electrolyte chemical potential, $\mu_{\rm el}$, is equal to that of the pure solid, $\mu_{\rm s}$, and (2) a ``direct coexistence approach" (DCA), in which the solution is equilibrated with a solid configuration (typically either a slab or a selected crystal environment) and the electrolyte concentration in the solution phase sufficiently far from the crystal surface is taken to be the solubility.  There also exist several implementation variants of the algorithms within each approach. We discuss the TA first.\\

Two basic TA variants have been used for calculating $m_s$.  The first, referred to here as TA1, calculates $\mu_{\rm el}$ at a number of concentration points, $m_i$ (algorithms for which are discussed in the next Section) in some neighborhood of the expected solubility value, fits them to an interpolation function, and then determines its crossing point with the solid chemical potential value, which is calculated separately.  The TA1 has been used by Vega \ea~\cite{Sanz2007,Aragones2012}, Paluch and Maginn~\cite{Paluch2010,Paluch2012}, and Panagiotopoulos \ea~\cite{Mester2015a,Mester2015b,Jiang2015}.  TA1 results can also be readily inferred from the $\mu_{\el}$ calculations of Mou\v{c}ka \ea~\cite{OEMC2011}, which we show later in this section.  \\

The TA2 variant implements an osmotic ensemble~\cite{OEMC2005,OEMC2011,OEMC2012,MNSGibbsDuhem2013,MNSLimitations2013,MNSActivity2015}
to directly calculate (in a single simulation run) the solubility, fixing the number of water molecules in the solution and setting $\mu(m)$ to the value $\mu_s$, with the resulting salt concentration yielding the solubility; note that this variant requires no chemical potential calculations in the solution phase.   Note also that by setting the imposed chemical potential to different values, the OEMC algorithm can also be used to calculate values of $\mu_{\rm el}(m)$ by calculating values of its inverse function, $m(\mu_{\rm el})$.    \\

In the interest of making relevant comparisons among the results of different workers in the following, some remarks are in order concerning the effects of different treatments of the electrostatics and the dependence of simulation results on system size. \\

Regarding the electrostatics, in an early implementation of the TA2 variant, Mou\v{c}ka \ea~\cite{OEMC2011} primarily used the Generalized Reaction Field (GRF) approach (some results obtained using Ewald summation were also presented), but in all subsequent papers the more conventional Ewald summation approach was used, in order that the results would be directly comparable with those of other researchers. An example of the effects of the different treatments on the calculated chemical potentials and solubility is indicated in the caption to Table \ref{solubilitydata}. \\

The notion of the system size dependence of a property conventionally means its dependence on the number of molecules in the simulation box for a given set of conditions, which for a mixture would refer to a given composition.  In the TA1 variant, the composition is set by using a fixed number of water molecules, $N_{\w}$, and varying numbers of salt molecules.  $N_{\w}=270$ has been previously used by Sanz and Vega~\cite{Sanz2007} and by Aragones \ea~\cite{Aragones2012}.  Panagiotopoulos \ea~\cite{Orozco2014,Mester2015a,Mester2015b,Jiang2015} generally used $N_{\w}=500$, except for concentrations below $m=0.56$ mol kg$^{-1}$.  The dependence of the results on the system size at a given concentration is studied by scaling the total number of molecules of both solvent and solute by the same factor.  Thus, Mester and Panagiotopoulos~\cite{Mester2015a} state that they studied systems of size 500 and 5000 water molecules at selected concentrations and found that the chemical potential results agreed to within their mutual simulation uncertainties. \\

A different issue regarding the system size arises intrinsically in the TA1 variant, in that the lowest concentration that can be studied is the case of a single NaCl molecule in $N_{\w}$ molecules, which requires large system sizes to access the low concentration region.  The lowest possible concentrations achievable for $N_{\w}=\{270,500,1000,5000\}$ are $m=\{0.206,0.111,0.056,0.011\}$, for which results were obtained by Mester and Panagiotopoulos~\cite{Mester2015a}.\\

The TA2 variant does not share this intrinsic system size limitation when accessing the low concentration region, and most simulations have been performed using $N_{\w}=270$.   The TA2 system size dependence has not been studied extensively, but Table I of reference \cite{Mester2015a} indicates agreement within the mutual simulation uncertainties of the TA2 results of Mou\v{c}ka \ea~\cite{Moucka2013} for the electrolyte chemical potential down to $m=0.01$ mol kg$^{-1}$, indicating that strong system size dependence of the TA2 implementation is unlikely.  Nevertheless, we believe that the system size dependence of results obtained using both the TA1 and TA2 methods at low concentrations should be further studied.
\\

Crystal properties obtained by different groups are summarized in Table \ref{solidmudata}, and the resulting solubilities calculated from simulations using the indicated $\mu_s$ values are given in Table \ref{solubilitydata}.  The former table shows that for a given FF and thermodynamic conditions there is generally good agreement with respect to the solid properties, but agreement with the mutually studied cases in the second column of Table \ref{solubilitydata} has only been achieved very recently.\\

Since it does not seem to be well--known in the simulation community, we briefly discuss the calculation of uncertainties in results obtained, either directly or indirectly, from more than one simulation quantity.  This arises both in the calculation of the solubility, considered here, and in the calculation of the electrolyte activity coefficient, discussed in Section \ref{section4.3}.\\

The uncertainty in $m_s$, $\delta m_s$, arises from the solution of the equation
\begin{equation}
\mu_{\rm el}(m) = \mu_s \label{solubilityequation}
\end{equation}
The TA1 and TA2 variants both require the value of $\mu_s$, the calculation of which has typically been performed by all groups using a variant of the Frenkel--Ladd Einstein crystal method~\cite{FrenkelLadd1984,Polson2000,Anwar2003,Vega2008} or the Einstein molecule method of Vega and Noya~\cite{Vega2007}.\\

At a given state point for a given set of simulation conditions (system size, length of simulation run and other simulation protocols), in the case of the TA1 variant, uncertainties in both $\mu_{\rm el}(m)$ and $\mu_s$ contribute to $\delta m_s$.  The contribution of the latter can be approximated by a Taylor expansion of Equation (\ref{solubilityequation}) about the calculated $m_s$ value, giving:
\begin{equation}
\delta \mu_{\rm sim} +
 \left(
 \frac{
 \partial \mu_{\rm el}
 }
 {\partial m}
 \right)_*
 \delta m_s
  = \delta \mu_s
 \label{sensitivity}
\end{equation}
where * denotes evaluation at the calculated $m_s$.  Combining the two contributions to $\delta m_s$ in the standard way (see, {\it e.g.}, \cite{nist2016,Smith2003a}) yields an estimated uncertainty (corresponding to one standard deviation) in the calculated solubility  of
\begin{equation}
\delta m_s =
\left [
\left(\delta \mu_s^2 + \delta \mu_{\rm sim}^2 \right)
/
\left(
\frac{\partial \mu_{\rm el}}{\partial m}
\right)_*^2
\right]^{1/2}  \label{sensitivityTA1}
\end{equation}

In the case of the TA2 (OEMC) variant, $\delta \mu_{\rm sim}$ is absent in Equation (\ref{sensitivity}), but that source of $\delta m_s$ must be augmented by the uncertainty $\delta m_{\rm sim}$ from the simulation run itself, giving
\begin{equation}
\delta m_s =
\left [
\delta \mu_s^2
/
\left(
\frac{\partial \mu_{\rm el}}{\partial m}
\right)_*^2 + \delta m_{\rm sim}^2
\right]^{1/2}  \label{sensitivityTA2}
\end{equation}

As an example, we consider the uncertainties of the calculated solubilities in Table \ref{solubilitydata} in light of Equations (\ref{sensitivityTA1}) and (\ref{sensitivityTA2}), in the case of the JC--SPC/E FF at ambient conditions.  We use a nominal value of 2.3 (in kJ kg mol$^{-2}$) for the term $(\partial \mu_{\rm el}/\partial m)_*$, obtained from our simulation data in Figure \ref{fig:FiguremuJC} (discussed later in the paper).  A value of $\delta \mu_s=0.2$ kJ mol$^{-1}$ is reasonably consistent with the scatter of the $\mu_s$ data in Table \ref{solidmudata}.  The value of $\delta \mu_{\rm sim}$ for the TA1 variant is reported by Mester \ea~\cite{Mester2015a} as 0.2 kJ mol$^{-1}$.  With these values, the TA1 solubility uncertainty is $\delta m_s=\sqrt{(0.2)^2 + (0.2)^2}/(2.3)=0.1$ mol kg$^{-1}$.  A typical value of $\delta m_{\rm sim}=0.2$ reported by Mou\u{cka \ea yields a TA2 solubility uncertainty $\delta m_s = \sqrt{(0.2/2.3)^2 + 0.2^2}= 0.2$ mol kg$^{-1}$. \\
}

Prior to the results of \cite{Mester2015a} and \cite{Vega2015}, an attempt to examine the reasons for the then different results was that of Mou\v{c}ka \ea~\cite{MNSGibbsDuhem2013}, who independently calculated both water and salt chemical potentials by the same TA2 osmotic algorithm and verified their mutual consistency with respect to the Gibbs-Duhem equation, thereby.  With a similar goal, Kobayashi \ea \cite{Kobayashi2014} carefully studied the details of the TA1 variant of the Vega group~\cite{Sanz2007,Aragones2012} and found that the calculated chemical potential values depend critically on the details of its implementation. \\

As indicated in Table \ref{solubilitydata}, the discrepancies among the different TA results have recently been resolved. The TA1 calculations of Mester and Panagiotopoulos~\cite{Mester2015a} and of Vega \ea~\cite{Vega2015} coincide with the earlier results of Mou\v{c}ka \ea~\cite{OEMC2011,MNSGibbsDuhem2013} for all FFs mutually considered.  It may thus be concluded that the TA produces correct results when it is correctly implemented.\\

The DCA also has several variants \cite{Alejandre2007,Alejandre2009,Alejandre2013,Manzanilla2015,Joung2009,Aragones2012,Kobayashi2014,Wiebe2015,Kolafa2015}, but solubility discrepancies among different implementations of the DCA and with the (correct) TA result have generally remained unresolved, with Table \ref{solubilitydata-DCA} indicating that most of the DCA solubilities are too high. The single exception is the very recent work of Kolafa \cite{Kolafa2015}, who studied a DCA variant
in which the (100) face of the crystal slab surface is replaced by a face with higher Miller indexes, {\it e.g.}, (145) or (257).  This appears to lower the activation energy required for the crystal to dissolve in a subsaturated solution environment or to grow in a supersaturated environment, and he has obtained a solubility value of about $m = 3.5$ mol kg$^{-1}$
for aqueous NaCl at ambient conditions, in good agreement with the most recent TA results~\cite{OEMC2011,Mester2015a,MNSGibbsDuhem2013,Vega2015}. Alejandr\'{e} \ea \cite{Manzanilla2015} search for a composition at which solid crystals are at the point of incipient formation, and other DCA researchers attempt to equilibrate the solution phase with a solid phase of some configuration. Possible difficulties with the implementation of the DCA are the length of the simulation run, the size and configuration of the solid phase used, and the total number of particles in the simulation.

\subsection{Electrolyte chemical potential}\label{section4.2}

Chemical potentials are arguably the most important solution properties, since all other thermodynamic properties may be calculated from their dependence on concentration, $T$ and $P$.  However, apart from their indirect use by means of a composition derivative in one of the Kirkwood--Buff integrals~\cite{Weerasinghe2003,Smith2010,Gee2011}, $\mu(m)$ at moderate and high concentrations has only recently been directly used within a TrS~\cite{MNSDevelopment2013}, but emphasis has instead been on the infinite dilution individual ion or total hydration free energies.  Although solution densities have been considered at moderate concentrations, they primarily provide information on the FF molecular size parameters.   A likely reason for the relative lack of use of higher concentration chemical potentials as TrS or TeS properties is that until recently their calculation has been deemed to be a very computationally intensive task.  This has been mitigated by the recent development of new algorithms discussed in this section, which include methods for calculating both electrolyte chemical potentials and solubility.\\

Since direct insertion of an electrolyte molecule using the standard Widom method~\cite{Widom1963} is computationally infeasible for aqueous electrolytes, a more subtle approach must be used for the calculation of $\mu_{\rm el}$, usually by means of some type of ``gradual particle insertion" method, in the form of either thermodynamic integration (TI)~\cite{Frenkel2002}, or an expanded ensemble (EE) approach ({\it e.g.}, ~\cite{Nezbeda1991}).   Both TI and EE methods express the electrolyte FF in terms of a selected function of a coupling parameter $\lambda$ (a ``$\lambda$--scaling function") that  yields a FF at $\lambda=0$ for a system whose properties are known (or easily calculated) and the full FF at $\lambda=1$.  They also require a choice of $\lambda$ values (corresponding to $\lambda$--states) over the interval [0,1], called a ``$\lambda$--staging strategy"; in its simplest form, this consists of equally spaced $\lambda$ values, with narrower spacing where rapid changes in the $\lambda$ dependence of relevant quantities occur.  For efficient implementation, an EE approach additionally requires a method to determine bias (balancing) factors associated with coupling parameter changes to ensure that the different $\lambda$ states are visited with roughly equal probability.\\

L\'{i}sal \ea~\cite{OEMC2005} were the first to consider both TI and EE  approaches for $\mu_{\rm el}$, and they employed several heuristic strategies for $\lambda$--scaling and $\lambda$--staging.  Their EE approach involved the use of MD simulations to generate molecular configurations (EE-MD algorithm).  They also proposed an osmotic ensemble method using MD (OEMD), which fixes the number of water molecules and externally imposes a chemical potential value, $\mu_{\rm el}$, to drive solute molecules into the simulation box and calculate their corresponding concentration, $m(\mu_{\rm el})$. \\

In a followup paper~\cite{OEMC2011}, Mou\v{c}ka \ea considered two extensions of the work in \cite{OEMC2005}. Their OEMC algorithm incorporates an improved method for implementing $\lambda$--scaling and $\lambda$--staging in conjunction with a self--adaptive Wang--Landau approach~\cite{Wang2001} to calculate biasing factors. We mention that the OEMC algorithm can calculate the electrolyte concentration corresponding to an arbitrary externally imposed chemical potential, thus calculating $\mu_{\rm el}(m)$ by means of the dependence of $m$ on $\mu_{\el}$, and can be easily adapted to consider chemical reactions~\cite{Smith1994,Turner2007} and to adsorption in nanoscale environments (see, {\it e.g.}, \cite{MouckaBratko2015}).\\

A multi--stage free energy perturbation (MFEP) algorithm for calculating $\mu_{\rm el}$ was also described in detail in the Appendix to \cite{OEMC2011}.  This algorithm generates molecular configurations using MC in conjunction with a strategy for implementing $\lambda$--scaling and $\lambda$--staging (MFEP-MC algorithm).  The approach of Panagiotopoulos \ea~\cite{Mester2015a,Mester2015b,Jiang2015}, implements a variation of this method.  They use MD to generate configurations in a MFEP--MD algorithm, which uses different $\lambda$--scaling and $\lambda$--staging strategies and the Bennett Acceptance Ratio method~\cite{Bennett1976} for the calculation of the chemical potential.\\

Shi and Maginn~\cite{ShiMaginn2007} had independently built on the research of \cite{OEMC2005} and of others to develop their Continuous Fractional Component Monte Carlo method.  This works in the context of an EEMC procedure in a canonical or osmotic ensemble, and they used it to calculate the phase behaviour of the CO$_2$--ethanol binary system by incorporating Wang--Landau sampling~\cite{Wang2001} for automatically determining bias factors.  In related work, Paluch \ea \cite{Paluch2010,Paluch2012} proposed a self--adaptive Wang--Landau transition matrix MC method within the context of an EE approach to efficiently calculate the solute chemical potential, and combined it with a pseudocritical path integration method for calculating the solid chemical potential.  They considered aqueous solutions of NaCl and of light alcohols.\\

Vega \ea~\cite{Sanz2007,Aragones2012,Kobayashi2014,Vega2015} proposed a different approach, using MD simulations to calculate $\mu_{\rm el}$ by means of a TI approach, which determines the difference between the total solution Gibbs function and that of a LJ pure--fluid reference system (whose properties are expressed analytically~\cite{Kolafa1994}).  Simulation data at a set of concentration points are fitted to an interpolating polynomial and $\mu_{\rm el}$ is obtained by differentiation with respect to particle number.\\

In the case of aqueous NaCl solutions (and similarly for other aqueous alkali metal halide solutions), the electrolyte chemical potential per mole of NaCl, $\mu_{\rm el} \equiv \mu_{\rm NaCl}$, at given $(T,P,m)$ is given by~\cite{OEMC2005,OEMC2011}:
\begin{eqnarray}
\mu_{\el} & = & \mu_{\rm Na^+} +  \mu_{\rm Cl^+} \nonumber \\
                        & = & \mu^{\rm IG}_{\el}+ \mu^{\rm ex}_{\el} \label{eq:mutotal}
\end{eqnarray}
where~\cite{OEMC2011}
\begin{eqnarray}
\mu^{\rm IG}_{\el} &   = &  \mu^0_{\rm Na^+}(T;P^0)+  \mu^0_{\rm Cl^-}(T;P^0)  + 2RT\ln \left(\frac{N_{\el}}{\beta P^0 <V>} \right) \label{eq:muIG}\\
\mu^{\rm ex}_{\el} & = & -RT\ln\left( \frac{\left<V^2<\exp(-\beta \Delta \cal U)>\right>}{<V^2>} \right) \label{eq:muex}
\end{eqnarray}
where $R$ is the universal gas constant, $\mu^0_{\rm Na^+}(T;P^0)+  \mu^0_{\rm Cl^+}(T;P^0)$ are the ion ideal--gas values at $T$ and the reference state pressure $P^0$ (available from standard thermochemical tables, {\it e.g.},\cite{JANAF1998}), and $<.>$ denotes a configurational average in the $NPT$ ensemble.\\

The MFEP methods in \cite{OEMC2011} and \cite{Mester2015a,Mester2015b,Jiang2015} evaluate the quantities in Equation (\ref{eq:muex}) by means of a $\lambda$--staging and $\lambda$--scaling strategy that can be viewed as calculating the ensemble average in the numerator of Equation (\ref{eq:muex}) by gradually inserting an electrolyte particle into the solution at a given concentration, and summing the chemical potential contributions over each $\lambda$ increment.  The OEMC algorithm essentially uses Equations (\ref{eq:mutotal}) to (\ref{eq:muex}) in a reverse manner, by fixing the number of water particles and $\mu_{\el}$, and calculating the corresponding equilibrium value of $N_{\rm NaCl}$ in the simulation. \\

\begin{figure} 
 \centering
 \hspace*{-3cm}\includegraphics[scale=1.0]{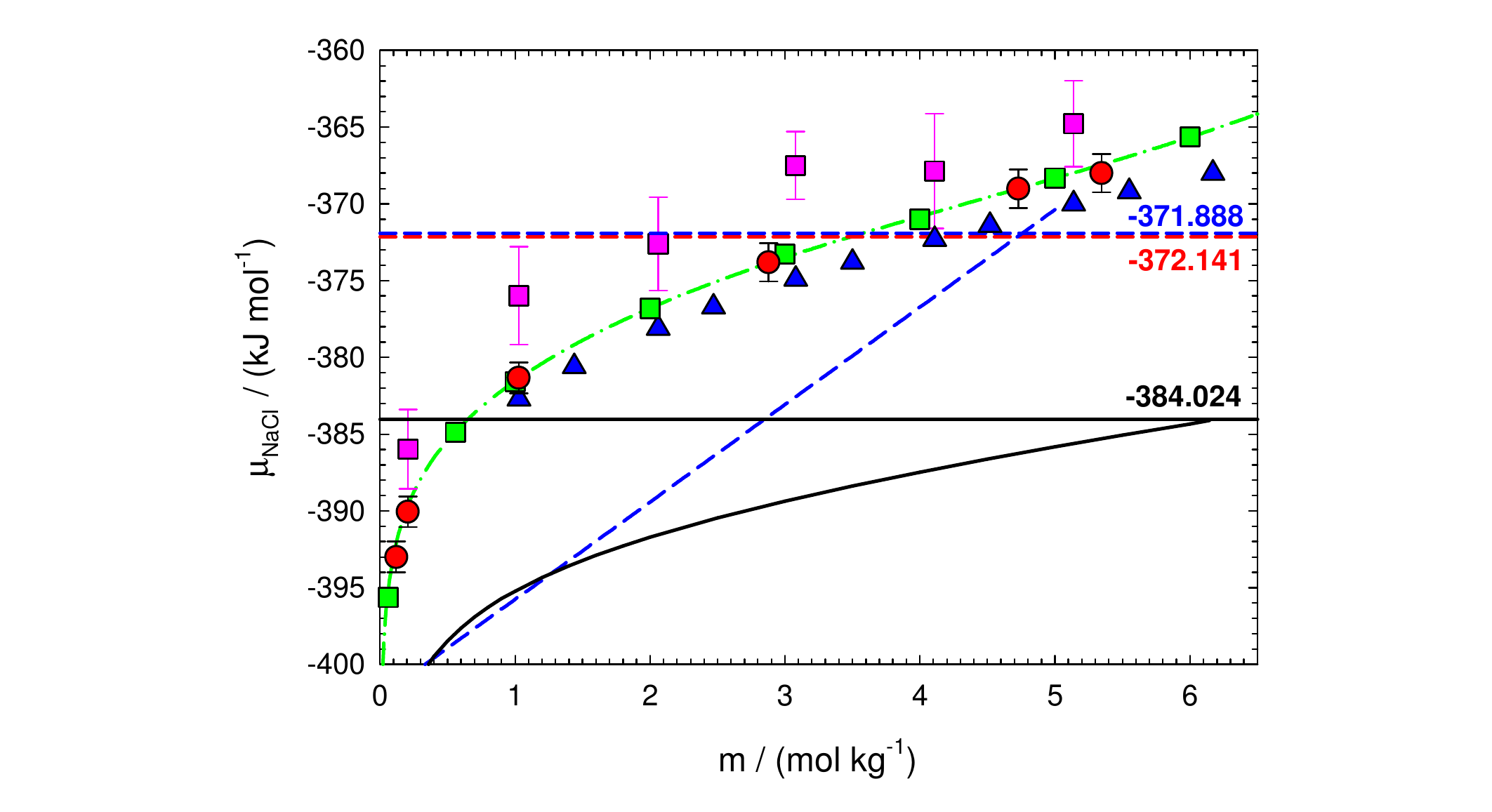}
 \caption{SD--BMHTF--SPC/E~\cite{SmithDang1994,Tosi1964} molar chemical potential, $\mu_{\rm NaCl}$, of NaCl in aqueous solution at ambient conditions ($T=298$~K or $T=298.15$~K and $P=1$~bar) as a function of molality, $m$. Symbols and curves correspond to data obtained by different authors and methodologies as follows: experimental data from Hamer and Wu at 298.15~K~\cite{Hamer1972} (black continuous curve), thermodynamic integration (TI) method of Sanz and Vega at 298~K~\cite{Sanz2007} (blue dashed curve), TI method with a higher precision of Aragones \ea at 298~K~\cite{Aragones2012} (blue triangles), MC transition matrix method of Paluch \ea \cite{Paluch2010,Paluch2012} at 298~K (magenta squares), MFEP--MC method of Mou\v{c}ka \ea at 298.15~K~\cite{OEMC2011} (red circles), MFEP--MD method of Mester and Panagiotopoulos at 298.15~K~\cite{Mester2015a} (green squares and dot-dashed curve). The black horizontal line denotes the experimental chemical potential value of crystalline NaCl \cite{JANAF1998} and the coloured dashed horizontal lines denote the values of the corresponding authors:
 $-371.88$~\cite{Aragones2012},
 -372.141~\cite{OEMC2011}.}
 \label{fig:FiguremuTF}
\end{figure}

Figure \ref{fig:FiguremuTF} compares results for $\mu_{\el}$ from different sources for the SD--BMHTF--SPC/E FF.  The MFEP--MC $\mu_{\el}$ simulation data of Mou\v{c}ka \ea~\cite{OEMC2011} agree with the MFEP--MD data of Mester and Panagiotopoulos~\cite{Mester2015a} within their mutual simulation uncertainties, and both lie well above the experimental curve. The results of Paluch \ea~\cite{Paluch2010} and of Aragones \ea~\cite{Aragones2012} lie respectively well above and slightly below these results. The results of Sanz and Vega~\cite{Sanz2007} are at variance with those of the other groups. We note that \cite{Aragones2012} used about 4 times longer simulation runs than in \cite{Sanz2007}.  The solubilities from \cite{OEMC2011} and \cite{Mester2015a} are $m\approxeq 3.6$ mol kg$^{-1}$, in mutual agreement within their simulation uncertainties. \\

\begin{figure} 
 \centering
 \hspace*{-3cm}\includegraphics[scale=1.0]{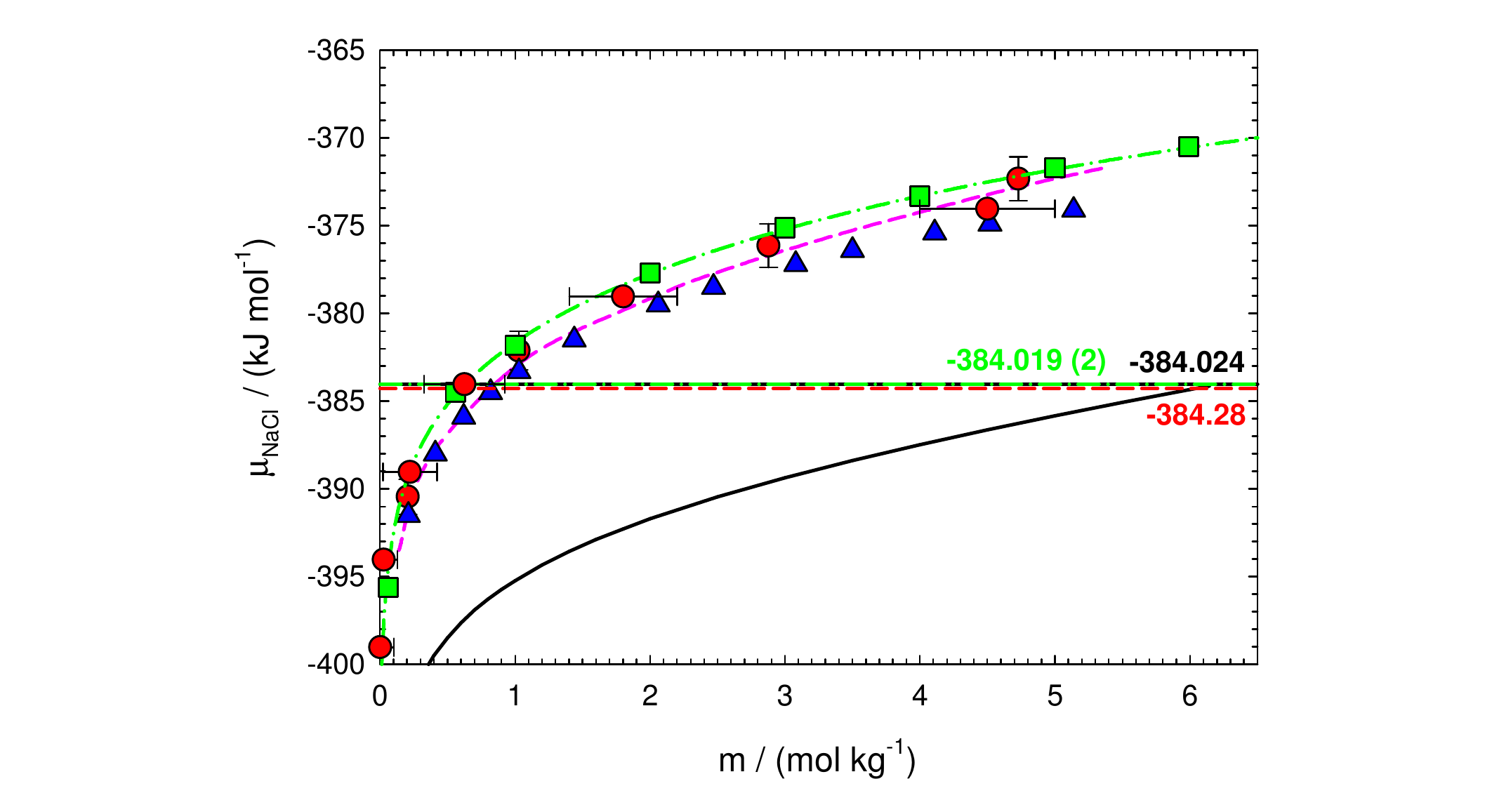}
 \caption{SD--SPC/E~\cite{SmithDang1994} molar chemical potential, $\mu_{\rm NaCl}$, of NaCl in aqueous solution at ambient conditions ($T=298$~K or $T=298.15$~K or $300$~K and $P=1$~bar) as a function of molality. Symbols and curves correspond to data obtained by different authors and methodologies as follows: Experimental data from Hamer and Wu \cite{Hamer1972} (black continuous curve), thermodynamic integration of Aragones \ea\cite{Aragones2012}
 (blue triangles), multi stage free energy perturbation of Mester and Panagiotopoulos \cite{Mester2015a} (green squares and green dash-dotted curve), averaged results using different methods of L\'{i}sal \ea \cite{OEMC2005} (magenta dashed curve), multi stage free energy perturbation of Mou\v{c}ka \ea \cite{OEMC2011} (red circles with vertical error bars), osmotic ensemble Monte Carlo of Mou\v{c}ka \ea \cite{MNSLimitations2013} (red circles with horizontal error bars). Also shown are corresponding chemical potential values of solid NaCl represented by the horizontal lines obtained by simulations of the same authors and experiment \cite{Wagman1982}.}
 \label{fig:FiguremuSD}
\end{figure}

Figure \ref{fig:FiguremuSD} compares results for $\mu_{\el}$ from different sources for the SD--SPC/E FF.   The MFEP--MC and OEMC results of Mou\v{c}ka \ea~\cite{OEMC2011} agree with the MFEP--MD results of Mester and Panagiotopoulos~\cite{Mester2015a} within their simulation uncertainties, and both lie well above the experimental curve.  The earlier results of L\'{\i}sal \ea obtained from several approaches~\cite{OEMC2005} are in general agreement with these results, and all 3 sets of results lie roughly parallel to the experimental curve at moderate and high concentrations. The results of Aragones \ea ~\cite{Aragones2012} are in slight disagreement with the others, particularly at higher concentrations. Although the solid chemical potential from this FF agrees well with the experimental value, since the $\mu_{\el}$ curves are much too high, the solubility is very low in comparison with experiment (see Table \ref{solubilitydata}).  \\

\begin{figure} 
 \centering
\hspace*{-3cm}\includegraphics[scale=1.0]{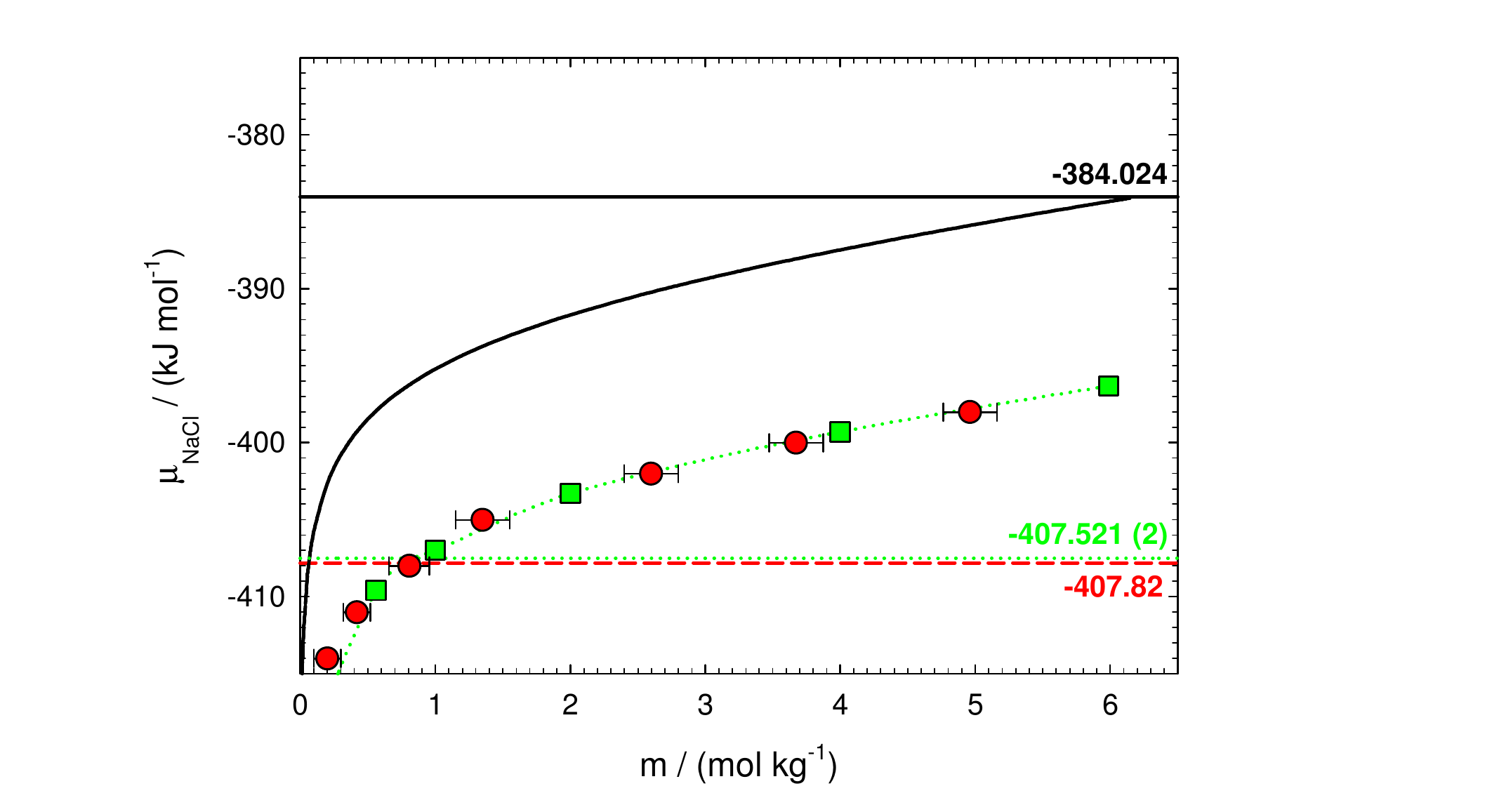}
 \caption{KBI--SPC/E~\cite{Weerasinghe2003} molar chemical potential, $\mu_{\rm NaCl}$, of NaCl in aqueous solution at ambient conditions ($T=298$~K or $T=298.15$~K or $300$~K and $P=1$~bar) as a function of molality. Symbols and curves correspond to data obtained by different authors and methodologies as follows: Experimental data from Hamer and Wu at 298.15~K \cite{Hamer1972} (black continuous curve), MFEP--MD results of Mester and Panagiotopoulos~\cite{Mester2015b} (green squares), OEMC results of Mou\v{c}ka \ea at 298.15~K~\cite{MNSLimitations2013} (red circles with horizontal error bars). The dotted green curve is an aid to the eye.  The black horizontal line denotes the experimental chemical potential value of crystalline NaCl in kJ mol$^{-1}$~\cite{JANAF1998} and the coloured horizontal lines denote the values of the corresponding authors: -407.82~\cite{MNSLimitations2013}, --407.521(2)~\cite{Mester2015b}.}
 \label{fig:FiguremuKBI}
\end{figure}

Figure \ref{fig:FiguremuKBI} shows results for $\mu_{\el}$ for the KBI--SPC/E FF from \cite{MNSLimitations2013} and \cite{Mester2015b} and their comparison with experiment.   The simulation results agree within their simulation uncertainties, and both lie well below and roughly parallel to the experimental curve at moderate and high concentrations.  Since the solid chemical potential of the FF is well below that of experiment, the solubility is very low (see Table \ref{solubilitydata}).\\

The upper panel of Figure \ref{fig:FiguremuJC} compares results for $\mu_{\el}$ from several sources for the JC--SPC/E FF. Again, the OEMC and the MFEP--MC results of Mou\v{c}ka \ea~\cite{OEMC2011} agree with the MFEP--MD results of Mester and Panagiotopoulos~\cite{Mester2015a} within their simulation uncertainties. All lie above the experimental curve, with increasing deviation at higher concentrations.  Since the solid chemical potential agrees well with the experimental value, the disagreement of the $\mu_{\el}$ and experimental curves at higher concentrations yield a solubility somewhat below the experimental result (see Table \ref{solubilitydata}). \\

Figure \ref{fig:FiguremuJCnew} shows results for $\mu_{\el}$ for the MNS-SPC/E FF of Mou\v{c}ka \ea~\cite{MNSDevelopment2013} in comparison with those of the JC--SPC/E FF and with experiment.  It is seen that, although the chemical potential curve of the former FF agrees better with experiment than the latter, its solid chemical potential is lower than the experimental value.  This results in the MNS--SPC/E FF  yielding a solubility values of $m=3.37(20)$ mol kg$^{-1}$, below the JC--SPC/E result of $m=3.64(20)$ mol kg$^{-1}$, and well below the experimental value of $m=6.14$ mol kg$^{-1}$.\\

Finally, the upper panel of Figure \ref{fig:FiguremuPolarizable} compares results for $\mu_{\el}$ from different sources for the polarizable FFs AH/BK3~\cite{AHBK32014} and AH/SWM4--DP~\cite{AHSWM4DP2006}, with the JC--SPC/E result and that of experiment shown for comparison.  As in the cases of the SD--SPC/E and KBI--SPC/E FFs, the AH/BK3 FF $\mu_{\el}$ curve is parallel to the experimental curve at moderate and high concentrations.  The AH/SWM4--DP results lie well below the experimental curve, and appear to terminate near $m=3$ mol kg$^{-1}$, indicating a limiting chemical potential and corresponding concentration similar to that shown in Figure \ref{fig:solubility-schematic}.  The AH/BK3 curve lies slightly below the experimental curve, in slightly better overall agreement with it than exhibited by the JC--SPC/E result.  However, its very low solid chemical potential value in comparison with experiment yields a solubility well below the experimental result (see Table \ref{solubilitydata}).

\subsection{Electrolyte activity coefficient}\label{section4.3}

The calculation of the electrolyte activity coefficient, $\gamma$, is directly related to the calculation of $\mu_{\el}$, and unrelated to the calculation of solubility.  Its value depends in general on the mathematical form of the model used to represent $\mu_{\rm el}$, which for aqueous electrolytes is conventionally expressed as~\cite{Hamer1972}:
\begin{equation}
 \mu(T,P,m) =   \mu^{\dag} (T,P) + 2RT \ln m + 2RT\ln \gamma(T,P,m) \label{MuHenry}
\end{equation}
where $\mu^{\dag}(T,P)$ is the chemical potential at the Henry-Law  reference state of a hypothetical 1 molal ideal solution and $\gamma$ is the corresponding activity coefficient.\\

Since both $\mu_{\el}$ on the left side of this equation and the ideal solution terms on the right side are typically very large, $\ln \gamma$ is the difference of two large quantities, both of which are subject to uncertainties. The details of the extreme sensitivity of $\ln \gamma$ calculations from simulation data to the value of  $\mu_{\el}^{\dag}$  will be illustrated later in this section.\\

Experimental data for $\ln \gamma$~\cite{Hamer1972} are available in terms of an empirical extension to the Debye--H\"{u}ckel theory~\cite{RobinsonStokes2002,Zemaitis1986}:
\begin{equation}
\ln \gamma =  \ln(10)\left(- \frac{A\sqrt{m}}{1 + B\sqrt{m}} + \beta m + c m^2 + dm^3\right) \label{eq:muefinal}
 \end{equation}
$A$ arises from the Debye-H\"{u}ckel theory, and is given by:
\begin{equation}
A = \frac{1.824 \times 10^6}{(\kappa T)^{3/2}} \label{eq:Aterm}
\end{equation}
where $\kappa$ is the dielectric permittivity of pure water for its FF at the solution ($T,P)$. The experimental value of $ \mu_{\el}^{\dagger}$ is available in thermochemical tables~\cite{Wagman1982} and values of the constants for various salt solutions at ambient conditions are given in \cite{Hamer1972}.\\

The determination of $\ln \gamma$ using Equation (\ref{MuHenry}) from the simulation values of $\mu_{\el}$ requires a method to determine $ \mu_{\el}^{\dag} (T,P)$.   Mou\v{c}ka \ea~\cite{MNSGibbsDuhem2013,MNSActivity2015} fitted their entire set of OEMC $\mu_{\el}$ simulation results to Equations (\ref{MuHenry}) and (\ref{eq:muefinal}), fixing $A$ and $B$ to the experimental values from Hamer and Wu~\cite{Hamer1972} and treating $(\mu^{\dag}_{\el},\beta,c,d)$ as fitting parameters.  Panagiotopoulos \ea~\cite{Mester2015a,Mester2015b,Jiang2015} used their MFEP--MD algorithm to calculate $\mu_{\el}$ and determined $\mu_{\el}^{\dag} (T,P)$ separately from their simulated value of $\mu_{\el}$ at the lowest concentration considered.  In \cite{Mester2015a}, they calculated $A$ from a simulation of $\kappa$ for the water FF in Equation (\ref{eq:Aterm}) and expressed $\ln \gamma$ in terms of the Debye--H\"{u}ckel low--concentration limit of Equation (\ref{eq:muefinal}):
\begin{equation}
\ln \gamma = -A\ln(10)\sqrt{m} \label{eq:Panag1}
\end{equation}
They used Equation (\ref{MuHenry}) to calculate $\mu^{\dag}$ at this low concentration value of $\mu_{\el}$ and then fitted the remainder of their remaining $\mu_{\el}$ simulation results to the parameters $(B, \beta,c,d)$.  In \cite{Mester2015b}, they proceeded similarly, but replaced (\ref{eq:Panag1}) by  the empirically based Davies equation~\cite{Davies1938}:
\begin{equation}
\ln \gamma = -A\left(\frac{\sqrt{m}}{1 + \sqrt{m}} - 0.2m \right)\ln(10) \label{eq:Panag2}
\end{equation}
They found in \cite{Mester2015b} that $\mu^{\dag}$ values obtained from using Equations (\ref{eq:Panag1}) and (\ref{eq:Panag2}) differed by about 0.01 kJ mol$^{-1}$, well within their simulation uncertainties.\\

The upper panel of Figure \ref{fig:BK3Gamma} shows the extreme sensitivity of $\ln \gamma$ to the value of $\mu^{\dag}$ using the OEMC simulation results of \cite{MNSActivity2015} for the AH/BK3 FF shown in the upper panel of Figure \ref{fig:FiguremuPolarizable}.  It shows the $\ln \gamma$ curves obtained by fixing $\mu^{\dag}_{\el}$ at the indicated values followed by fitting $(\beta,c,d)$ to the entire set of $\mu_{\el}$ simulation results in the upper panel of Figure \ref{fig:FiguremuPolarizable} while fixing $A$ and $B$ to the experimental values; the lower panel shows the resulting $\mu_{\el}$ curves for the indicated values of $\mu^{\dag}$.  $\mu^{\dag}=-398.793$ is the result obtained in \cite{MNSActivity2015}. $\mu^{\dag}=-397.0(3)$ is the value reported by Jiang \ea~\cite{Jiang2015} in the analysis of their MFEP--MD simulation data, which they found to yield excellent agreement with the experimental $\ln \gamma$ curve within their simulation uncertainties.  Figure \ref{fig:BK3Gamma} shows that
$\mu^{\dag}=-396.480$ yields a $\ln \gamma$ curve in perfect coincidence with the experimental curve in Figure \ref{fig:BK3Gamma}, while leaving the $\mu_{\el}$ curve in the lower panel essentially unchanged on the scale of the graph. (We do not recommend this approach to calculate an accurate value of $\mu^{\dag}$; see instead Equation (\ref{eq:mudagger}) below. The upper panel of Figure 10 is intended solely to emphasize the fact that the $\ln \gamma$ curve is extremely sensitive to the $\mu^{\dag}$ value.)
\\

The ability shown in the upper panel of Figure \ref{fig:BK3Gamma} to precisely fit the experimental $\ln \gamma$ curve at moderate and high concentrations by adjusting the value of $\mu^{\dag}$ is a consequence of the fact that the AH/BK3 $\mu_{\el}$ curve in the lower panel is parallel to the corresponding experimental curve at moderate and high concentrations, and shifted from it by a nearly constant amount.  This is a general result for a FF displaying such behaviour, which can be seen by considering $\mu_{\rm el}$ in Equation (\ref{MuHenry}) for the simulation and experimental curves at a given concentration within this range:
\begin{eqnarray}
\mu_{\rm sim} & = & \mu_{\rm sim}^{\dag} + 2RT\ln m + 2RT\ln \gamma_{\rm sim}\\
\mu_{\rm expt} & = & \mu_{\rm expt}^{\dag} + 2RT\ln m + 2RT\ln \gamma_{\rm expt}
\end{eqnarray}
Subtraction gives
\begin{equation}
\mu_{\rm sim} - \mu_{\rm expt} = \mu_{\rm sim}^{\dag} - \mu_{\rm expt}^{\dag} + 2RT\ln \left(\frac{\gamma_{\rm sim}}{\gamma_{\rm expt}}\right)
\end{equation}
If $\mu_{\rm sim} - \mu_{\rm expt}$ is independent of concentration, then agreement of $\gamma_{\rm sim}$ with $\gamma_{\rm expt}$ can be accomplished by setting
\begin{equation}
\mu_{\rm sim}^{\dag} = \mu_{\rm expt}^{\dag} + [\mu_{\rm sim} - \mu_{\rm expt}]
\end{equation}
Thus, when a $\mu_{\el}$ curve at moderate and high concentrations is parallel to the experimental curve, agreement of its $\ln \gamma$ curve with experiment can be achieved by selecting an appropriate $\mu_{\rm el}^{\dag}$ value. However, this may not agree with the experimental result, which in the case of NaCl at ambient conditions is -393.133 kJ mol$^{-1}$~\cite{Wagman1982}.\\

In view of the strong dependence of $\ln \gamma_{\rm sim}$ on the value of $\mu^{\dagger}_{\rm sim}$ and the different values for NaCl obtained thus far by essentially indirect approaches, in addition to its intrinsic importance as a basic quantity in the Henry--Law chemical potential model, we recommend that its value is best determined for a given FF by means of a separate calculation.
This may be achieved by equating the chemical potential expressions of Equations (\ref{eq:mutotal})--(\ref{eq:muex}) and (\ref{MuHenry}) and taking the infinite dilution limit, where $\ln \gamma=0$ in Equation (\ref{MuHenry}), to yield
the following new expression that allows direct comparison with experimental values ({\it e.g.}, ~\cite{Wagman1982}).
\begin{equation}
\mu_{\rm NaCl}^{\dag}(T,P)= \mu_{\rm Na^+}^{0}(T;P^0)+ \mu_{\rm Cl^-}^{0}(T;P^0)+2 RT\ln \left(\frac{RTM_{\w}}{v_{\w}(T,P)P^0}\right) +  \mu_{\rm NaCl}^{ex;\infty} \label{eq:mudagger}
\end{equation}
where $M_{\w}$ is the molecular weight of H$_2$O in kg mol$^{-1}$ and $v_{\w}$ is its molar volume in m$^3$ mol$^{-1}$, and $\mu_{\rm NaCl}^{ex;\infty}$ is the excess chemical potential in Equation (\ref{eq:muex}) at infinite dilution. The uncertainty in $\mu_{\rm NaCl}^{\dag}(T,P)$ using this expression is given by the simulation uncertainty in $\mu_{\rm NaCl}^{ex;\infty}$.  Calculations using Equation (\ref{eq:mudagger}) are currently being studied by our group.\\

The uncertainty $\delta \ln \gamma$ associated with a MFEP method such as those in references
\cite{OEMC2011} and \cite{Jiang2015} is given approximately from Equation (\ref{MuHenry}) by
\begin{equation}
\delta \ln \gamma = \frac{1}{2RT}\left[(\delta \mu)^2 + (\delta \mu^{\dag})^2\right]^{1/2} \label{eq:uncertainty1}
\end{equation}
where $\delta \mu$ and $\delta \mu^{\dag}$ are the uncertainties in $\mu$ and $\mu^{\dag}$, respectively. For example, in the MFEP--MD method of \cite{Jiang2015} at 298.15~K, $\delta \mu \approxeq \delta \mu^{\dag} \approxeq 0.3$, yielding $\delta \ln \gamma \approxeq 0.1$.
The corresponding uncertainty in $\ln \gamma$ associated with the OEMC method (which we remind the reader was originally developed for a different purpose) is given approximately by
\begin{equation}
\delta \ln \gamma = \left[\left(\frac{\delta \mu^{\dag}}{2RT}\right)^2  +\left(\frac{\delta m}{m}\right)^2 \right]^{1/2} \label{eq:uncertainty2}
\end{equation}
Using $\delta \mu^{\dag}=0.3$ from \cite{Jiang2015} and $\delta m = 0.2$ from \cite{MNSActivity2015} at a typical value of $m=0.4$ yields $\delta \ln \gamma \approxeq 0.5$, which arises primarily from the uncertainty in $\ln m$.  Thus, to be competitive with the MFEP approach for calculating $\ln \gamma$, the OEMC method requires that its uncertainty in $\ln m$ be similar to the uncertainty in $\mu/2RT$ in the former approach, which in the study of \cite{Jiang2015} at 298.15~K is about 0.06. This would require simulation runs about 10 times longer than were used in \cite{MNSActivity2015}.  Thus, in the following comparisons we will only display $\ln \gamma$ results from the MFEP--MD approach of Panagiotopoulos \ea~\cite{Mester2015a,Mester2015b,Jiang2015}.\\

Figure \ref{fig:ln-gamma-mester} compares results for $\ln \gamma$ for several FFs with those of experiment; the simulation data points are the MFEP--MD simulation data of the Panagiotopoulos group~\cite{Mester2015a,Mester2015b,Jiang2015}. The KBI--SPC/E and AH/BK3 results are seen to fall on the experimental curve, and the results for the other FFs deviate from it to varying degrees.  In contrast to this excellent agreement for $\ln \gamma$, Figures \ref{fig:FiguremuKBI} and \ref{fig:FiguremuPolarizable} show that the KBI--SPC/E and AH/BK3 $\mu_{\rm el}$ results are roughly parallel to but shifted by a constant amount from the experimental curve at moderate and high concentrations (the shift is relatively small for AH/BK3 but quite large for KBI--SPC/E).  \\

Examination of Figure \ref{fig:FiguremuSD} for the SD--SPC/E FF shows that its $\mu_{\el}$ curve also displays similar behaviour for the $\mu_{\el}$ curve at moderate and high concentrations, resulting in the relatively good agreement with experiment of the $\ln \gamma$ values in Figure \ref{fig:ln-gamma-mester}, and we conjecture that it can be brought into exact coincidence by a slightly altered value of $\mu_{\rm el}^{\dag}$.\\

Finally, we remark that the KBI approach uses the concentration derivative of $\mu_{\rm el}$ at a moderately high concentration as a TrS property, which is essentially equivalent to using the concentration derivative of its $\ln \gamma$ for this purpose.  In order for the $\ln \gamma$ curve itself to agree with experiment, additional information is required. Weerasinghe and Smith in \cite{Weerasinghe2003} imposed the known experimental low--concentration behaviour of the $\ln \gamma$ derivative in order to achieve the good agreement with experiment for the entire $\ln \gamma$ curve shown in Figure 4 of \cite{Weerasinghe2003}.\\

\subsection{Water chemical potential and related properties}\label{section4.4}

We close by discussing the water chemical potential, $\mu_{\w}$, and some properties related to it that have been of recent interest in the literature.\\

$\mu_{\w}$ can be expressed experimentally as
\begin{equation}
\mu_{\rm H_2O} = \mu^*_{\rm H_2O}(T,P) + RT \ln a_{\rm H_2O}(T,P,m)
\end{equation}
where
$\mu^*_{\rm H_2O}(T,P)$ is the chemical potential of pure water at the system $(T,P)$ and $a_{\rm H_2O}$ is its activity.  $\mu_{\rm H_2O}$ may either be calculated in separate simulations or obtained from $\mu_{\w}$ by integrating the Gibbs-Duhem equation using Equations (\ref{MuHenry}) and (\ref{eq:muefinal}), yielding \cite{MNSGibbsDuhem2013}
\begin{eqnarray}
\mu_{\w}(T,P,m) & = & \mu_{\w}^*(T,P) -2RT m M_{\w} -RTM_{\w}\ln(10) \nonumber \\
                & & \times\left(\beta m^2 + \frac{4}{3}Cm^3 + \frac{3}{2}Dm^4 + \frac{2A}{B^3 + B^4\sqrt{m}} \right. \nonumber \\
                & & \left. + \frac{4A\ln(B\sqrt{m} + 1)}{B^3}- \frac{2A\sqrt{m}}{B^2}-\frac{2A}{B^3}\right) \label{eq:muw}
\end{eqnarray}
Similarly to the requirement for a value of $\mu^{\dag}$ in Equation (\ref{MuHenry}), a value for $\mu_{\w}^*(T,P)$ is required in Equation (\ref{eq:muw}), which may either be calculated in a separate simulation or by treating it as one of the parameters in fitting Equation (\ref{eq:muw}) to $\mu_{\w}$ simulation data.  Equations (\ref{MuHenry}) and (\ref{eq:muw}) may also be used to test the consistency with respect to the Gibbs--Duhem equation of separately calculated values of $\mu_{\el}$ and $\mu_{\w}$, as was done by Mou\v{c}ka \ea in \cite{MNSGibbsDuhem2013}.\\

\begin{figure} 
\centering
 \hspace*{-3cm}\includegraphics[scale=1.0]{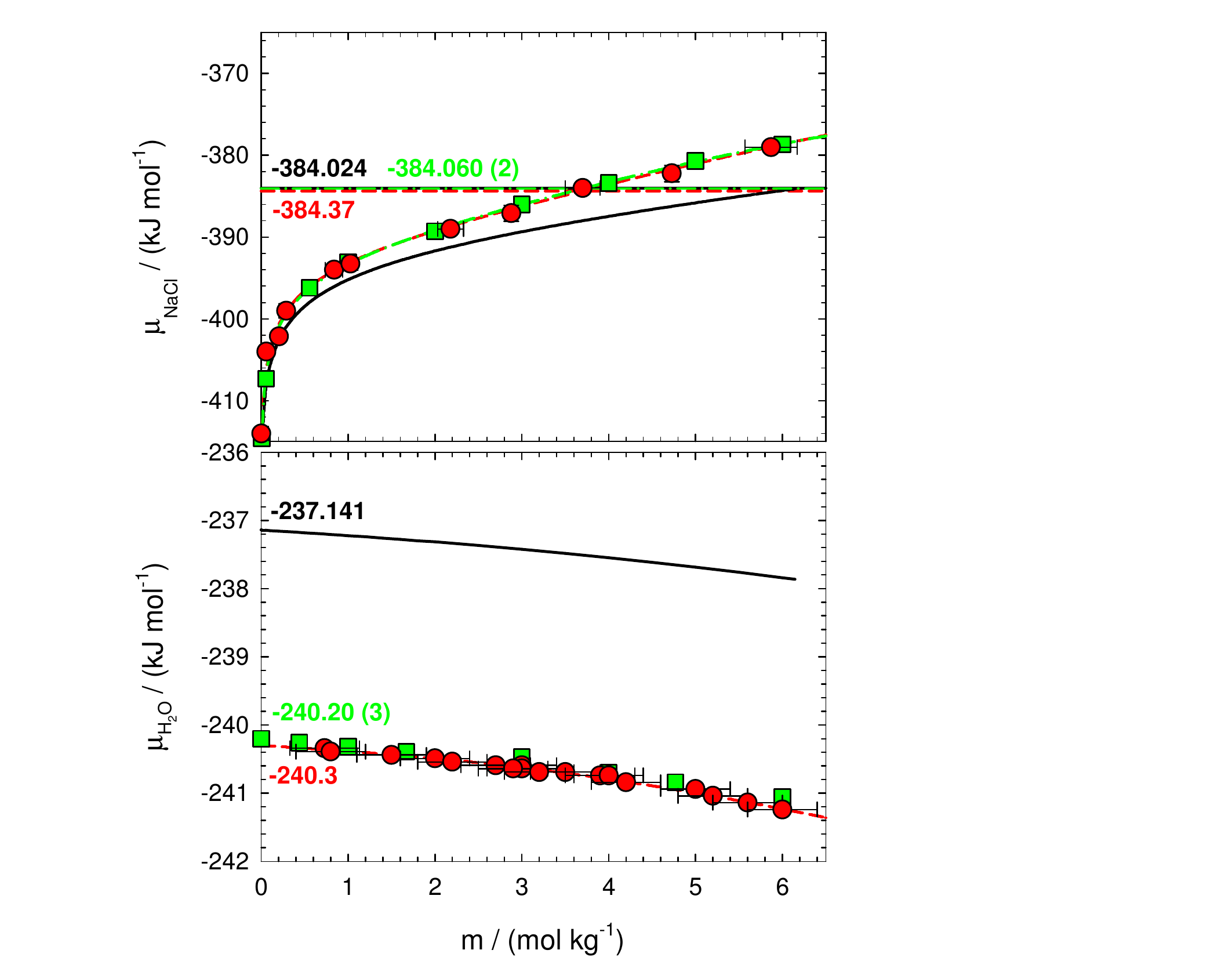}
 \caption{The upper panel shows the JC--SPC/E~\cite{Joung2008} molar chemical potential, $\mu_{\rm NaCl}$, of NaCl in aqueous solution at $T=298.15$~K and $P=1$~bar as a function of molality, $m$. Symbols and curves correspond to data obtained by different authors and methodologies as follows: experimental data from Hamer and Wu \cite{Hamer1972} (black continuous curve), multi--stage free energy perturbation algorithm of Mou\v{c}ka \ea \cite{OEMC2011} (red circles with vertical error bars), OEMC method of Mou\v{c}ka \ea \cite{MNSGibbsDuhem2013} (red circles with horizontal error bars and red dashed curve), MFEP--MD of Mester and Panagiotopoulos~\cite{Mester2015a} (green squares and green dash-dotted curve). The black horizontal line denotes the experimental chemical potential value of crystalline NaCl~\cite{JANAF1998} and the coloured dashed horizontal lines denote the values of the corresponding authors: -384.37~\cite{MNSLimitations2013,MNSDevelopment2013}, -384.060(2)~\cite{Mester2015a}.  The lower panel shows the corresponding water chemical potential curves, using the same colour coding as in the upper panel.  The pure water chemical potentials are from the following sources: -237.141~\cite{JANAF1998}, -240.3~\cite{MNSLimitations2013,MNSDevelopment2013}, -240.20(3)~\cite{Mester2015a}.}
 \label{fig:FiguremuJC}
\end{figure}

\begin{figure} 
 \centering
 \hspace*{-3cm}\includegraphics[scale=1.0]{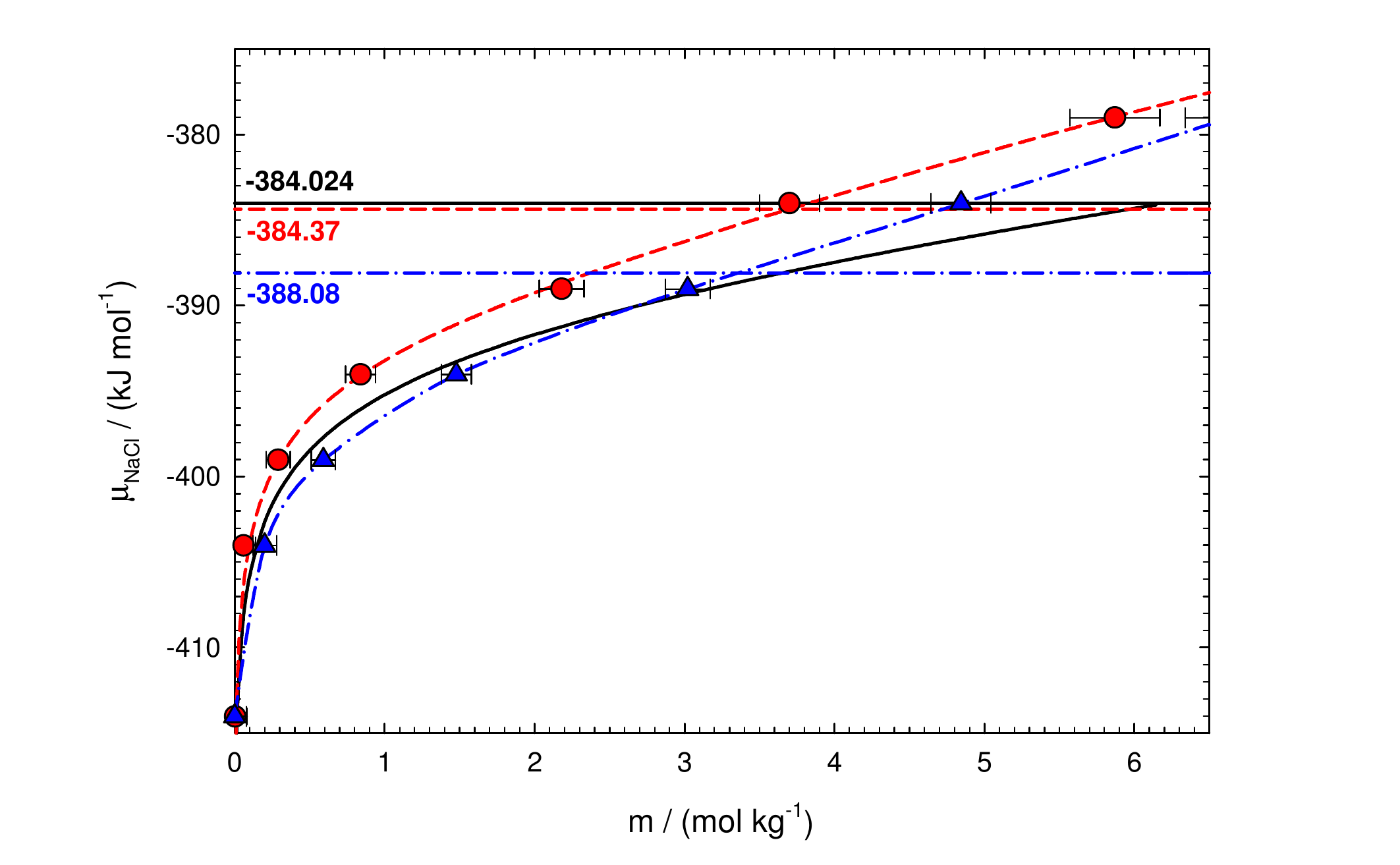}
 \caption{JC--SPC/E~\cite{Joung2008} chemical potential, $\mu_{\rm NaCl}$, of NaCl in aqueous solution at $ T=298.15$~K and $P=1$~bar as a function of molality, $m$, obtained by the OEMC method \cite{MNSGibbsDuhem2013} (red circles and red dashed curve) and for the improved force field of Mou\v{c}ka \ea \cite{MNSDevelopment2013} (blue triangles and blue dash-dotted curve) compared to the experimental data from Hamer and Wu \cite{Hamer1972} (black continuous curve). The horizontal lines denote the corresponding chemical potential values of crystalline NaCl: Experiment \cite{JANAF1998} (black continuous line), Frenkel-Ladd method of Mou\v{c}ka \ea \cite{MNSLimitations2013} (the other two lines).}
 \label{fig:FiguremuJCnew}
\end{figure}

\begin{figure} 
 \centering
\hspace*{-3cm}\includegraphics[scale=1.0]{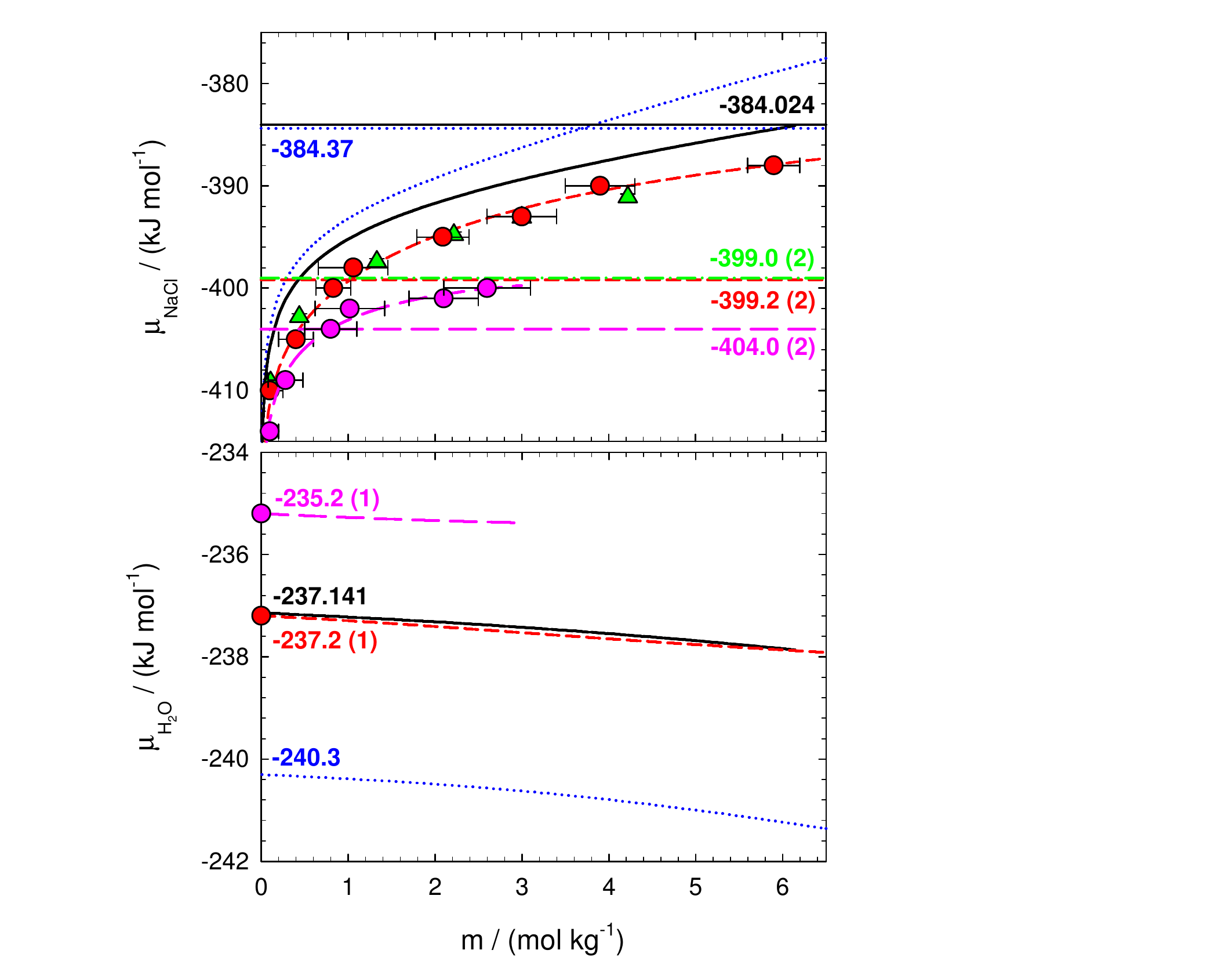}
 \caption{AH/BK3~\cite{AHBK32014} and AH/SWM4--DP~\cite{AHSWM4DP2006} molar chemical potential of NaCl (upper panel) and of water (lower panel) in aqueous solution at $ T=298.15$~K and $P=1$~bar as a function of molality, $m$, by different methods and authors: OEMC method of Mou\v{c}ka \ea \cite{MNSActivity2015} for the AH/BK3 FF (red circles and red dashed curves), OEMC method of Mou\v{c}ka \ea \cite{MNSActivity2015} for the AH/SWM4--DP FF (magenta circles and red dashed curves), MFEP--MD method of Jiang \ea \cite{Jiang2015} for the AH/BK3 FF (green triangles). The horizontal lines denote corresponding chemical potential values of crystalline NaCl: Frenkel--Ladd method of Mou\v{c}ka \ea \cite{MNSActivity2015} (red dashed line at -399.2 kJ mol$^{-1}$), Frenkel--Ladd method of Mou\v{c}ka \ea \cite{MNSActivity2015} (magent dashed line at -404.0(2) kJ mol$^{-1}$, Frenkel--Ladd method of Jiang \ea \cite{Jiang2015} (green dash-dotted line at -399.0(2) kJ mol$^{-1}$). For comparison, also shown are the results of Mou\v{c}ka \ea for the nonpolarizable SPC/E and Joung and Cheatham force field \cite{MNSGibbsDuhem2013} (blue dotted curves and solid chemical potential of -384.37 kJ mol$^{-1}$) and the experimental data (black continuous curves \cite{Hamer1972} and solid chemical potential of -384.024~\cite{JANAF1998}.}
 \label{fig:FiguremuPolarizable}
\end{figure}

\begin{figure} 
 \centering
\hspace*{-3cm}\includegraphics[scale=1.0]{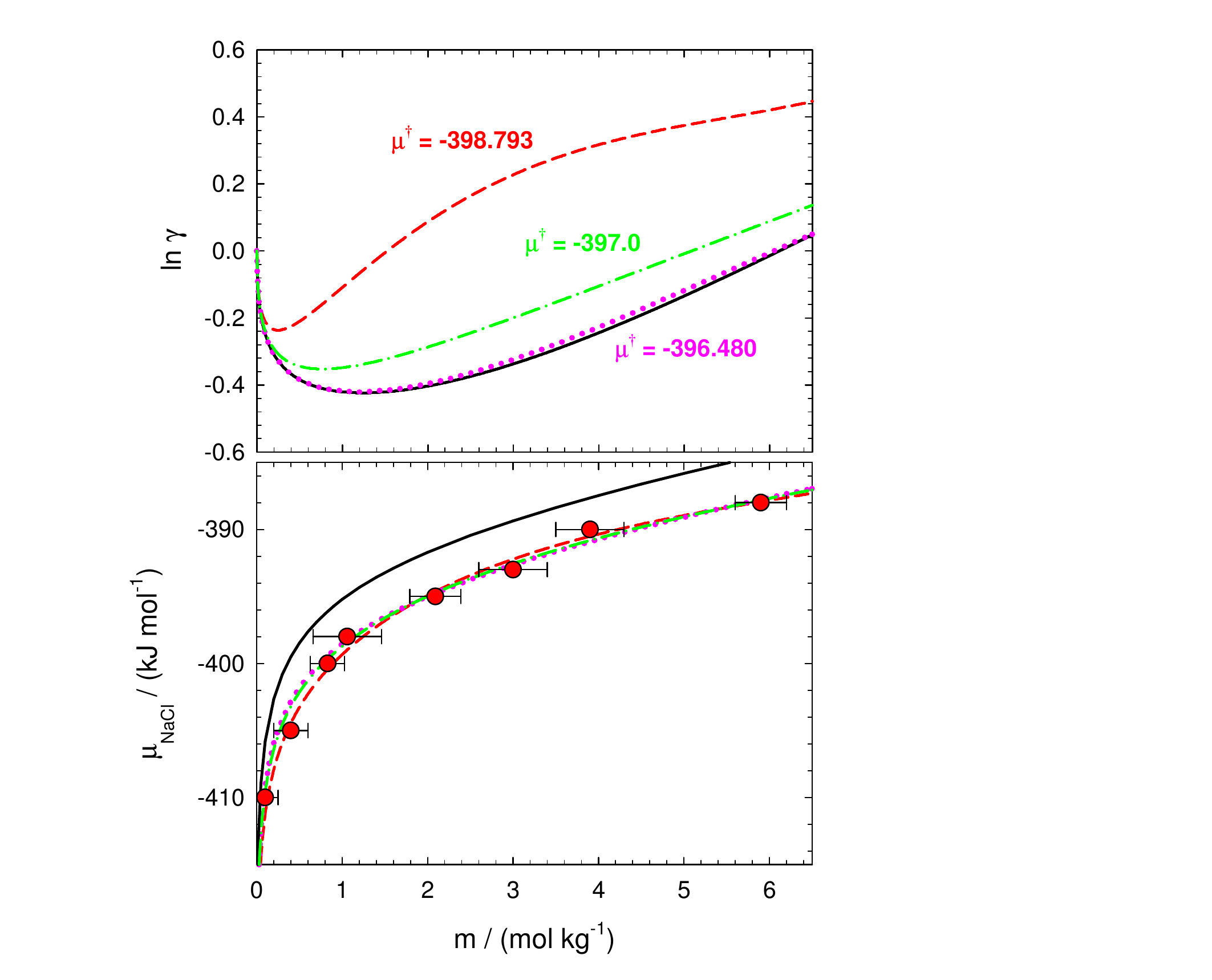}
 \caption{Upper Panel: Concentration dependence of the natural logarithm of the NaCl Henry convention molal activity coefficient, $\ln \gamma$, for the AH/BK3 \cite{AHBK32014} force field and its comparison with the experimental results \cite{Hamer1972}. The red dashed curve was obtained by considering the standard state chemical potential, $\mu^{\dagger}$, as a free fitting parameter \cite{MNSActivity2015}. The magenta dotted curve was obtained by fitting to simulation data at the fixed value of $\mu^{\dagger}=-396.48$~kJ~mol$^{-1}$ \cite{MNSActivity2015}, and the green dash-dotted curve at the fixed value $\mu^{\dagger}=-397.0$~kJ~mol$^{-1}$. The latter value was taken from Jiang \ea \cite{Jiang2015}. The black continuous curve is the experimental result taken from Hamer and Wu \cite{Hamer1972}.\\
Lower Panel: Corresponding concentration dependence of the
molar NaCl chemical potential,
$\mu_{\el}$ \cite{MNSActivity2015}. The points are the simulation results and their uncertainties (one standard deviation) are indicated.}
 \label{fig:BK3Gamma}
\end{figure}

\begin{figure} 
 \centering
\hspace*{-3cm}\includegraphics[scale=1.0]{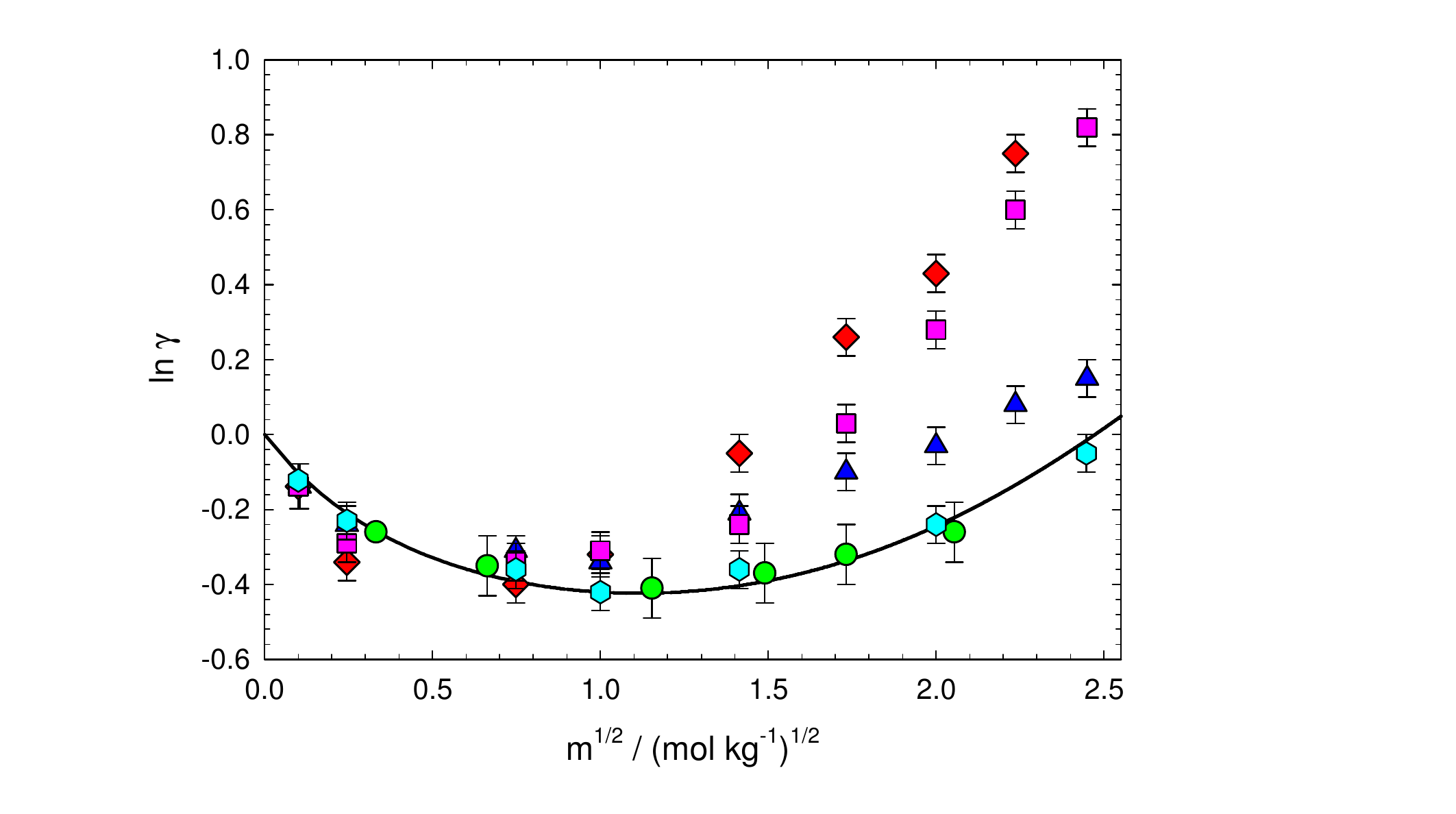}
 \caption{Predicted concentration dependence of the natural logarithm of the NaCl Henry convention molal activity coefficient, $\ln \gamma$, for several force fields, from Mester and Panagiotopoulos \cite{Mester2015a}:AH/BK3 (green circles), SD--SPC/E (blue triangles), JC--SPC/E (magenta squares), SD--BMHTF--SPC/E (red diamonds), KBI--SPC/E (cyan hexagons). Experimental curve (black) is from Hamer and Wu [113].}
 \label{fig:ln-gamma-mester}
\end{figure}

The curves in the lower panels of Figures \ref{fig:FiguremuJC} and \ref{fig:FiguremuPolarizable} show $\mu_{\rm H_2O}$ for the JC--SPC/E and AH/BK3 FFs, respectively, obtained in each case by integrating the Gibbs-Duhem equation using the simulation values of $\mu_{\el}$ fitted to Equations (\ref{eq:muefinal}) and (\ref{MuHenry}).  In Figure \ref{fig:FiguremuJC}, the $\mu_{\w}$ curve of Mou\v{c}ka \ea~\cite{MNSGibbsDuhem2013} and the simulation data of Mou\v{c}ka \ea~\cite{MNSGibbsDuhem2013} and of Mester and Panagiotopoulos~\cite{Mester2015a} are shown for comparison, along with the experimental curve.  The simulation curves agree perfectly on the scale of the graph, with the value for $\mu^*_{\w}$ of the former being slightly lower than that of the latter.  However, their agreement with the experimental curve is poor, due to the neglect of polarizability by the JC--SPC/E FF.   The lower panel of Figure \ref{fig:FiguremuPolarizable} shows a similar comparison of $\mu_{\w}$ for the AH/BK3 and SWM4--DP FFs from \cite{MNSActivity2015} with experiment and with the JC--SPC/E results.  It is seen that the AH/BK3 result is in excellent agreement with the experimental curve, whereas the AH/SWM4--DP results lie well above those of experiment.\\

The osmotic pressure, $\Pi$, and the vapor pressure, $P^*(T,P,m)$, of aqueous NaCl solutions are directly related to the water activity in the solution, due to the requirement for the equality of its chemical potential.\\

W.R. Smith \ea~\cite{Smith2015a} determined $\Pi$ from the relevant thermodynamic relations for the water chemical potential and pure water phases.  $\Pi$ is given by
\begin{equation}
\Pi(T,P,m) =  -\frac{RT \ln a_{\w}(T,P,m)}{\overline{v}_{\w}(T,P,m)}
\end{equation}
where $\overline{v}_{\w}(T,P,m)$ is the partial molar volume of water in the solution phase, which can be calculated from simulation values of the concentration dependence of the solution density (and can be well approximated at ambient conditions by the pure water molar volume).  They calculated $a_{\w}$ using Equation (\ref{eq:muw}) with its parameter values obtained by fitting the simulation results for $\mu_{\rm NaCl}$ to Equation (\ref{eq:muw}) with $\gamma$ given by Equation (\ref{MuHenry}).  They considered an aqueous NaCl solution phase at ambient conditions for the JC--SPC/E, AH/BK3 and SWM4--DP FFs, and found that AH/BK3 gave the best agreement with experiment.   \\

We remark that a MD method to calculate the osmotic pressure, $\Pi$, was recently developed by Luo \ea~\cite{Luo2010,Luo2013}, which has similarities to earlier work of Murad \ea~\cite{Murad1993,Murad1995,Powles1995,Paritosh1996}.  Its concentration dependence was used by them and later by Saxena and Garcia~\cite{Saxena2015} as a TrS property for the determination of FF parameters.  This is essentially equivalent to using the water activity for this purpose.\\

The solution vapor pressure, $P^*(T,P,m)$ can also be calculated from the water activity and partial molar volume by means of thermodynamic relations.  Orozco \ea~\cite{Orozco2014} calculated $P^*(T,P,m)$ using Gibbs Ensemble MC simulations for several FFs and temperatures.  They found that this approach at 298K had relatively large statistical uncertainties, and no results were shown for the JC--SPC/E FF.  \\

\begin{figure} 
 \centering
\hspace*{-3cm}\includegraphics[scale=1.0]{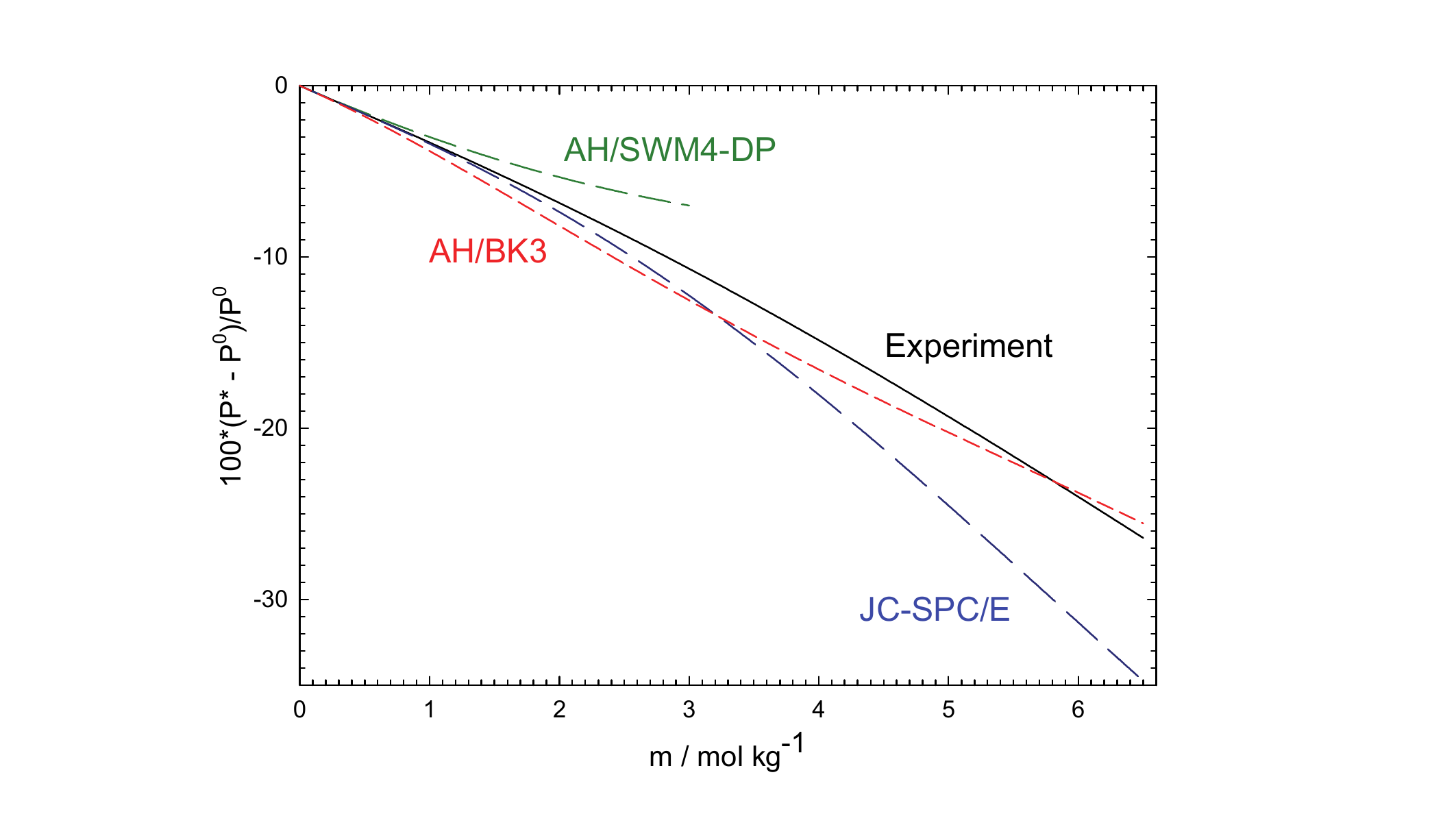}
 \caption{
 Concentration dependence of the relative decrease in vapour pressure of an aqueous NaCl solution as a function of molality, $m$, for the indicated force fields; from Smith \ea \cite{Smith2015b}.
 The pure water vapour pressures of the respective FFs \{JC--SPC/E~\cite{Joung2008}, AH/BK3~\cite{AHBK32014}, AH/SWM4--DP~\cite{AHSWM4DP2006}\} and Experiment are (all units in kPa) $\{0.88, 3.09, 6.95\}$ and 3.17~\cite{REFPROP}.}
 \label{fig:Pstar}
\end{figure}

Here, we consider instead some preliminary results~\cite{Smith2015b} (a full account will be published elsewhere) for the calculation of $P^*$ from the water chemical potential simulation results using a thermodynamic approach.
Equating $\mu_{\w}$ in the vapor and solution phases, and relating the chemical potentials and partial molar volumes of H$_2$O to their values at a pressure $\overline{P}$ at which they are available gives
\begin{eqnarray}
RT\ln a_{\w}(T,\overline{P},m) + (P^* - \overline{P})\overline{v}_{\w}(T,\overline{P},m)&  = & (\widehat{P}^*-\overline{P})\overline{v}_{\w}(T,\overline{P},0) \nonumber\\
& & + RT \ln \left( \frac{P^*}{\widehat{P}^*}\right) \nonumber \\
& & + (P^*-\widehat{P}^*)B(T) \label{Pvap}
\end{eqnarray}
where $\widehat{P}^*$ is the vapor pressure of the pure water FF at $T$ and $B(T)$ is the second virial coefficient of
the pure water FF
at the temperature $T$. This equation may be solved numerically for $P^*$, and some results are shown in Figure \ref{fig:Pstar}
at $T=298.15$ K using $B_2=0$, which should be a reasonable approximation in view of the low absolute vapour pressures.  As can be seen, the AH/BK3 FF produces the best overall results. The JC--SPC/E results are slightly better than the AH/BK3 results up to about $m=0.35$ mol kg$^{-1}$, but deteriorate rapidly at higher concentrations.  The AH/SWM4--DP results are poor.

\section{Summary and Recommendations}\label{section5}

The previous sections have presented recent results concerning force field development and the calculation of chemical potentials, solubility and activity coefficients for aqueous electrolytes, with a focus on the simplest case of aqueous NaCl solutions at ambient conditions.  In the lists below, we summarize our findings and make suggestions for future progress, arranged by topic.
\subsection{FF development}\label{section5.1}
\begin{enumerate}
\item
Polarizable force fields are preferred over nonpolarizable force fields.
\item
The incorporation of a vdW site on the H atoms may improve predictions for the electrolyte solution properties; this is in accord with a suggestion of Alejandr\'{e} \ea~\cite{Alejandre2009}.
\item
Special treatment of the cation--O interaction may improve the electrolyte FF; the KBI approach of P. Smith \ea~\cite{Weerasinghe2003,Smith2010,Gee2011} used an empirical modification of the geometric mean combining rule.
\item
In the determination of FF parameters, the set of properties used to fit the parameters (TrS properties) should be explicitly distinguished from properties predicted from the resulting FF (TeS properties).
\end{enumerate}
\subsection{Training set (TrS) properties}\label{section5.2}
\begin{enumerate}
\item The following aqueous electrolyte thermodynamic properties should be considered for inclusion in the TrS: \begin{itemize}
    \item density at moderately high concentration
    \item electrolyte chemical potential at both low and moderate concentrations
    \item solid density and chemical potential
    \item electrolyte solubility
    \item ion--water structure
 \end{itemize}
    Note that certain other potential TrS properties are subsumed by those in the above list ({\it e.g.}, the KBI related to the concentration derivative of the electrolyte chemical potential curve is a subsidiary property of its chemical potential, and the osmotic pressure is a subsidiary property of the water chemical potential, which in turn is related to the electrolyte chemical potential via the Gibbs--Duhem equation).
\item Single--ion hydration properties must be used with care (see the discussion in Section \ref{section3.1}.
\end{enumerate}
\subsection{Solubility}\label{section5.3}
\begin{enumerate}
\item 
Thermodynamic--based approaches for calculating the electrolyte solubility produce mutually consistent results when they are appropriately implemented.
\item 
Further studies are required to ascertain why,
apart from the recent results of Kolafa~\cite{Kolafa2015} that agree with the thermodynamic approaches, other current implementations  of the Direct Coexistence Approach (DCA) for calculating electrolyte solubility do not yield the correct result.
\item 
 The TA2 osmotic ensemble method (OEMC or OEMD) is more computationally efficient than a
 TA1 approach for calculating solubility,
 since it can be implemented in a single simulation run.  It can be also used for multi--electrolyte solutions, where a TA1 variant would be impractical.
\item 
Uncertainties should be given in published simulation results whenever feasible.  These are given for the solubilities from the TA1 and TA2 approaches in Equations (\ref{sensitivityTA1}) and (\ref{sensitivityTA2}), respectively.
\end{enumerate}
\subsection{Chemical potentials}\label{section5.4}
\begin{enumerate}
\item 
The calculation of $\mu_{\rm el}(m)$ is best carried out by means of
the multi--stage free energy perturbation (MFEP) approach, via either the MFEP--MC~\cite{OEMC2005,OEMC2011} or the MFEP--MD~\cite{Mester2015a,Mester2015b,Jiang2015} method,
or the OEMC method~~\cite{OEMC2005,OEMC2011,OEMC2012,MNSGibbsDuhem2013,MNSLimitations2013,MNSActivity2015}.
\item 
The OEMC approach for the calculation of electrolyte chemical potentials
is easily extended to the cases of aqueous multi--electrolyte solutions, chemically reacting solutions and adsorption in nanoscale environments.
\item 
The calculation of the water chemical potential may be an alternative means of calculating all chemical potentials in an aqueous solution of a single solute if it is easier to simulate than the solute chemical potential, which may be subsequently calculated from the Gibbs--Duhem equation. However, at least one point on the $\mu_{\rm el}$ curve must be still obtained separately as the Gibbs-Duhem equation provides only the dependence between the derivatives of the two curves. This approach can also be used to test the self consistency of the algorithms implemented for each chemical potential.
\item 
The system size dependence of $\mu_{\el}(m)$ simulation results obtained by both TA1 and TA2 approaches should be investigated further, particularly at low concentrations and near the limiting concentration in Figure \ref{fig:solubility-schematic}.
\end{enumerate}
\subsection{Activity coefficients}
\begin{enumerate}
\item 
$\delta \ln \gamma$, the uncertainty in the calculated value of $\ln \gamma$ is unrelated to the uncertainty in the calculation of the solubility, apart from a common contribution of the uncertainty in the calculation of the chemical potential.  $\delta \ln \gamma$ is given by Equations (\ref{eq:uncertainty1}) and (\ref{eq:uncertainty2}) for the TA1 and TA2 approaches, respectively.
\item 
$\ln \gamma$ should be determined from Equation (\ref{MuHenry}) using smoothed results for $\mu_{\el}$ (for example, fitted to Equations (\ref{eq:muefinal}) and (\ref{MuHenry}),
in conjunction with
an accurate value of $\mu^{\dag}$.
\item 
$\mu^{\dag}$, in Equation (\ref{MuHenry}) should be calculated as accurately as possible, and separately from any numerical manipulation of the set of $\mu_{\rm el}(m)$ simulation results.
We have presented a new method for such a calculation in Equation (\ref{eq:mudagger}).
\item 
The uncertainties in $\ln \gamma$ from the MFEP and OEMC methods are given by Equations (\ref{eq:uncertainty1}) and (\ref{eq:uncertainty2}), respectively.
\end{enumerate}

\section{Acknowledgements}

Support for this work was provided by the Natural Sciences and Engineering Research Council of Canada (Discovery Grant OGP1041), the SHARCNET (Shared Hierarchical Academic Research Computing Network) HPC
consortium (http://www.sharcnet.ca), and the Czech National Science Foundation (Grant No. P106-15-19542S). Access to computing and storage facilities owned by parties and projects contributing to the Czech National Grid Infrastructure MetaCentrum, provided under the program ``Projects of Large Infrastructure for Research, Development, and Innovations" (LM2010005), is greatly appreciated.


\clearpage
\begin{landscape}
\begin{table}
    \centering
    \caption{NaCl crystal chemical potentials, $\mu_s$ used in the calculation of solubilities in the corresponding rows of Table \ref{solubilitydata}, and densities, $\rho_s$, at $P=1$~bar for several force fields from different research groups.  All results are at $T=298.15$~K except those of Vega {\it et al.}, which are at $T=298$~K. All results were obtained using 500 ion pairs except those indicated by *, which were obtained by extrapolation to a system of infinite size.  The values of chemical potentials are given in kJ~mol$^{-1}$ of NaCl and the values of density in kg~m$^{-3}$.  The experimental values are $\mu_s=-384.024$ kJ mol$^{-1}$~\cite{JANAF1998} and $\rho_s=2165$ kg m$^{-3}$~\cite{Hamer1972}. Unless indicated otherwise, the reference refers to the original source.}
\bigskip
    \begin{tabular}{lclllc}
    \hline
    \hline
    Force Field   & Property & W.R. Smith \ea                                             & \multicolumn{2}{c}{Panagiotopoulos \ea }                       & Vega \ea \\
    \hline
   SD--SPC/E & $\mu_s$ & -384.28\cite{MNSLimitations2013}	&	 value from \cite{MNSLimitations2013}        &	 \multicolumn{1}{c}{$-$}	 &	-384.07~\cite{Aragones2012}\\
             &         &      	&	-383.764\cite{Mester2015b}	        &	 -384.019(2)*\cite{Jiang2015}	&	 \multicolumn{1}{c}{$-$}\\
             & $\rho_s$ & 1932\cite{MNSLimitations2013}	&	 1931.65(7)\cite{Mester2015b}	        &	 1932.4(1)*\cite{Jiang2015}	&	 1932.~\cite{Aragones2012}\\
   \hline
   KBI--SPC/E & $\mu_s$ & -407.82\cite{MNSLimitations2013}	&	 -407.220\cite{Mester2015b}	        &	 -407.521(2)*\cite{Mester2015b}    &	\multicolumn{1}{c}{$-$}\\
              & $\rho_s$ & 2112\cite{MNSLimitations2013}	&	 2108.33(4)\cite{Mester2015b}	        &	 2109.5(1)*\cite{Mester2015b}    &	 \multicolumn{1}{c}{$-$}\\
   \hline
   SD--BMHTF & $\mu_s$ & 	  -372.141~\cite{OEMC2012}   &	value from \cite{Aragones2012}  &\multicolumn{1}{c}{$-$}	   &	 -372.10$^{\dag}$~\cite{Sanz2007}\\
   --SPC/E & & \multicolumn{1}{c}{$-$}	         &	 \multicolumn{1}{c}{$-$}	                                        &\multicolumn{1}{c}{$-$}	   &	-371.88~\cite{Aragones2012}\\
    & $\rho_s$ & 	2116.99 (this work)     &	\multicolumn{1}{c}{$-$}	                                        &    \multicolumn{1}{c}{$-$}                                    &	 2125.~\cite{Sanz2007}\\
    &  & \multicolumn{1}{c}{$-$}     &	\multicolumn{1}{c}{$-$}	                                        &    \multicolumn{1}{c}{$-$}                                    &	 2116.~\cite{Aragones2012}\\
   \hline
   JC--SPC/E & $\mu_s$   & -384.37\cite{MNSDevelopment2013}	&	value from \cite{MNSDevelopment2013}   &	 	   \multicolumn{1}{c}{$-$} &	-384.12~\cite{Aragones2012}\\
             &         &  -384.05(2) (this work) 	&	 -383.946 \cite{Mester2015b}	        &	 -384.06(2)*\cite{Mester2015b}	   &	value from \cite{Aragones2012}\\
             &    $\rho_s$  & 2010.88(4)\cite{MNSDevelopment2013}	&	 2010.38(6) \cite{Mester2015b}	    &	 2010.7(1)*\cite{Mester2015b}	   &	 2010.~\cite{Aragones2012}\\
   \hline
   MNS--SPC/E & $\mu_s$   & -388.08\cite{MNSDevelopment2013}	&	 \multicolumn{1}{c}{$-$}	        &\multicolumn{1}{c}{$-$}	 &	 \multicolumn{1}{c}{$-$}\\
             &    $\rho_s$  & 1978.61(4)\cite{MNSDevelopment2013}	&	 \multicolumn{1}{c}{$-$} 	    &	\multicolumn{1}{c}{$-$}	   &	\multicolumn{1}{c}{$-$}\\
   \hline
   AH/SWM4--DP & $\mu_s$ & -404.0(2)\cite{MNSActivity2015}   &      \multicolumn{1}{c}{$-$}                                       &   \multicolumn{1}{c}{$-$}                                      &  \multicolumn{1}{c}{$-$} \\
              & $\rho_s$ &2095.9 (this work)   &   \multicolumn{1}{c}{$-$}                                          & \multicolumn{1}{c}{$-$}                                        & \multicolumn{1}{c}{$-$}  \\
   \hline
   AH/BK3    & $\mu_s$        & -399.2(2)\cite{MNSActivity2015}      &  \multicolumn{1}{c}{$-$}                                           & -399.0(2)*\cite{Jiang2015}       & \multicolumn{1}{c}{$-$}\\
                & $\rho_s$ & 2116.6 (this work)\cite{MNSActivity2015}      &  \multicolumn{1}{c}{$-$}                                           & 2115.8(4)*\cite{Jiang2015}       & \multicolumn{1}{c}{$-$}\\
	\hline
    \end{tabular}
	\label{solidmudata}
\end{table}
\noindent $^{\dag}$ Obtained in \cite{OEMC2012} from the original value of -758.9~kJ~mol$^{-1}$ in \cite{Sanz2007} by transformation to the reference state used in this work.
\end{landscape}

\clearpage

\begin{table}
    \centering
    \caption{NaCl aqueous solubilities, $m_s$, in mol kg$^{-1}$ H$_2$O, calculated using the Thermodynamic Approach (TA) at $P=1$~bar for several force fields by different research groups from the values of $\mu_s$ in the corresponding rows of Table \ref{solidmudata}.
        All results use Ewald summation for the electrostatics and hence may be compared within the rows of the table delineated by solid lines.
    All results are at $T=298.15$~K except those of Vega {\it et al.}, which are at $T=298$~K.
    When there is more than a single value in a given row, results indicated in italics are in mutual agreement.}
\bigskip
    \begin{tabular}{lccrc}
    \hline
    \hline
    Force Field   & W.R. Smith \ea                                             & Panagiotopoulos \ea                        & Vega \ea & Paluch \ea\\
    \hline
SD--SPC/E  & {\it 0.61}\cite{MNSLimitations2013}  &	 {\it 0.61(1)}\cite{Mester2015a} &	0.9(4)\cite{Aragones2012}  & \multicolumn{1}{c}{$-$}\\
           & \multicolumn{1}{c}{$-$}              &	{\it 0.63(1)}*\cite{Mester2015b} &
           {\it 0.63(20)}\cite{Vega2015}\\
\hline
KBI--SPC/E & {\it 0.83}\cite{MNSLimitations2013}	        &	 {\it 0.88(2)}*\cite{Mester2015b}	        &\multicolumn{1}{c}{$-$} & \multicolumn{1}{c}{$-$}	    \\
\hline
SD--BMHTF  & {\it 3.6(2)}$^{\dag}$	            &	 {\it 3.57(5)}\cite{Mester2015a}          &	   	 5.4(8)~\cite{Sanz2007} & 0.8(2)~\cite{Paluch2010}\\
--SPC/E                & \multicolumn{1}{c}{$-$}         &	 \multicolumn{1}{c}{$-$}   	& 4.3(3)~\cite{Aragones2012} & \multicolumn{1}{c}{$-$}\\
\hline
   JC--SPC/E   & {\it 3.64(20)}\cite{MNSLimitations2013}  & {\it 3.59(4)}\cite{Mester2015a} 	
               & 4.8(3)~\cite{Aragones2012}        & \multicolumn{1}{c}{$-$}\\
               &
{\it 4.0(2)}$^{\ddag}$
          & {\it 3.71(4)}*\cite{Mester2015b}
               & {\it 3.71(20)}\cite{Vega2015} & \multicolumn{1}{c}{$-$}\\
\hline
 MNS--SPC/E    & 3.37(20)\cite{MNSDevelopment2013}       &  \multicolumn{1}{c}{$-$}                              & \multicolumn{1}{c}{$-$} & \multicolumn{1}{c}{$-$}              \\
\hline
   AH/SWM4--DP  & 0.8(2)\cite{MNSActivity2015}       &  \multicolumn{1}{c}{$-$}                              & \multicolumn{1}{c}{$-$} & \multicolumn{1}{c}{$-$}              \\
\hline
AH/BK3         & {\it 1.0(2)}~\cite{MNSActivity2015}   & {\it 0.99(5)}*~\cite{Jiang2015}   & \multicolumn{1}{c}{$-$} & \multicolumn{1}{c}{$-$}    \\
\hline
    \end{tabular}
	\label{solubilitydata}
\end{table}
\noindent Results in column 2 were obtained by the OEMC (TA2) except as indicated, and the others by TA2.\\
\noindent * Obtained using $\mu_s$ in Table \ref{solidmudata} calculated by extrapolation to infinite system size\\
\noindent $^{\dag}$ Estimated from the MFEP--MC Ewald summation $\mu_{\el}(m)$ curve of Figure 6 (open triangles) of \cite{OEMC2011} and our
value $\mu_s=-372.141$ kJ mol$^{-1}$
in Table \ref{solidmudata}.  The Generalized Reacton Field (GRF) curves in this figure give a solubility of 4.3(2) mol kg $^{-1}$.\\
\noindent $^{\ddag}$ Estimated from the MFEP--MC Ewald summation $\mu_{\el}(m)$ curve of Figure 6 (open circles) of \cite{OEMC2011} and our
value $\mu_s=-384.05$ kJ mol$^{-1}$
in Table \ref{solidmudata}.  The GRF curves in this figure give a solubility of 4.8(2) mol kg $^{-1}$.

\clearpage
\begin{table}
    \centering
    \caption{NaCl aqueous solubility, $m_s$, in mol kg$^{-1}$ H$_2$O, calculated using the Direct Coexistence Approach (DCA) at $P=1$~bar for the JC--SPC/E force field by different research groups.}
\bigskip
    \begin{tabular}{llc}
    \hline
    \hline
    Authors                             & \multicolumn{1}{c}{Method}                 & $m_s$\\
    \hline
    Joung and Cheatham~\cite{Joung2009}, 298~K & slab geometry       & 7.27(7) \\
    Aragones \ea~\cite{Aragones2012}, 298~K    & slab geometry       & 5.5(4)\\
    Kobayashi \ea~\cite{Kobayashi2014}, 298~K  & slab geometry, small system size      & 8.07\\
    Kobayashi \ea~\cite{Kobayashi2014}, 298~K  & slab geometry, large system size       & 6.20(23)\\
    Wiebe \ea~\cite{Wiebe2015}, 300~K          & seed crystal        &  5.7\\
    Manzanilla--Granados \ea~\cite{Manzanilla2015}, 298.15~K & slab geometry & 6.0(4)\\
                                                   & seed crystal  & 5.8(2)\\
    Kolafa~\cite{Kolafa2015}, 298.15~K             &
    higher Miller indices than (100)
    & 3.5\\
    \hline
\end{tabular}
	\label{solubilitydata-DCA}
\end{table}

\clearpage

\clearpage

\clearpage

\clearpage

\end{document}